\documentclass[oneeqnum,onefignum,onetabnum,onethmnum]{siamltex}
\usepackage{graphics}

\newcommand{\bc}{\begin{center}}
\newcommand{\ec}{\end{center}}
\newcommand{\beq}{\begin{equation}}
\newcommand{\eeq}{\end{equation}}
\newcommand{\bea}{\begin{eqnarray}}
\newcommand{\eea}{\end{eqnarray}}

\def\e23{\epsilon^{2/3}}
\def\cb{\bar{c}}

\def\Eb{{\bar{E}}}

\def\ct{\tilde{c}}
\def\rhot{{\tilde{\rho}}}
\def\Et{{\tilde{E}}}

\def\Ea{{\acute{E}}}

\def\cc{\check{c}}

\def\Ec{{\check{E}}}
\def\phic{{\check{\phi}}}
\def\cbr{\breve{c}}
\def\rhobr{{\breve{\rho}}}
\def\Ebr{{\breve{E}}}

\def\Eg{{\grave{E}}}

\def\i{j}
\def\ir{\i_r}

\title{Electrochemical thin films at and above the classical limiting current}

\author{
Kevin T. Chu\footnotemark[2]
\and 
Martin Z. Bazant\footnotemark[2]}

\date{ June 12, 2004 } 

\begin{document}
\maketitle

\renewcommand{\thefootnote}{\fnsymbol{footnote}}
\footnotetext[2]{Department of Mathematics, Massachusetts Institute of
Technology, Cambridge, MA 02139}
\renewcommand{\thefootnote}{\arabic{footnote}}

\begin{abstract}
We study a model electrochemical thin film at dc currents exceeding
the classical diffusion-limited value.  The mathematical problem
involves the steady Poisson-Nernst-Planck equations for a binary
electrolyte with nonlinear boundary conditions for reaction kinetics
and Stern-layer capacitance, as well as an integral constraint on the
number of anions.  At the limiting current, we find a nested boundary
layer structure at the cathode, which is required by the reaction
boundary condition.  Above the limiting current, a depletion of anions
generally characterizes the cathode side of the cell.  In this regime,
we derive leading-order asymptotic approximations for the (i)
classical bulk space-charge layer and (ii) another, nested highly
charged boundary layer at the cathode. The former involves an exact
solution to the Nernst-Planck equations for a single, unscreened ionic
species, which may apply more generally to Faradaic conduction through
very thin insulating films. By matching expansions, we derive
current-voltage relations well into the space-charge regime.
Throughout our analysis, we emphasize the strong influence of the
Stern-layer capacitance on cell behavior.
\end{abstract}



\pagestyle{myheadings}
\thispagestyle{plain}
\markboth{KEVIN T. CHU AND MARTIN Z. BAZANT}
{ELECTROCHEMICAL FILMS ABOVE THE LIMITING CURRENT}

\section*{Introduction} 
Thin-film technologies offer a promising way to construct \\
rechargeable micro-batteries, which can be directly integrated into 
modern electronic
circuits~\cite{dudney1995,wang1996,neudecker2000,takami2002,shi2003,nava}.
Due to the power-density requirements of many applications, such as
portable electronics, micro-batteries are likely to be operated under
at high current density, possibly exceeding diffusion limitation. In a
thin film, very large electric fields are easily produced by applying
only small voltages, due to the small electrode separation, which may
be comparable to the Debye screening length.  Under such conditions,
the traditional postulates of macroscopic electrochemical
systems~\cite{newman_book,rubinstein_book} --- bulk electroneutrality
and equilibrium double layers --- break down near the classical
diffusion-limited current~\cite{part1}. The mathematical justification
for these postulates is based on matched asymptotic expansions in the
limit of thin double
layers~\cite{chernenko1963,newman1965,macgillivray1969},
which require subtle modifications at large currents.

\begin{figure}
\bc
\scalebox{0.35}{\includegraphics{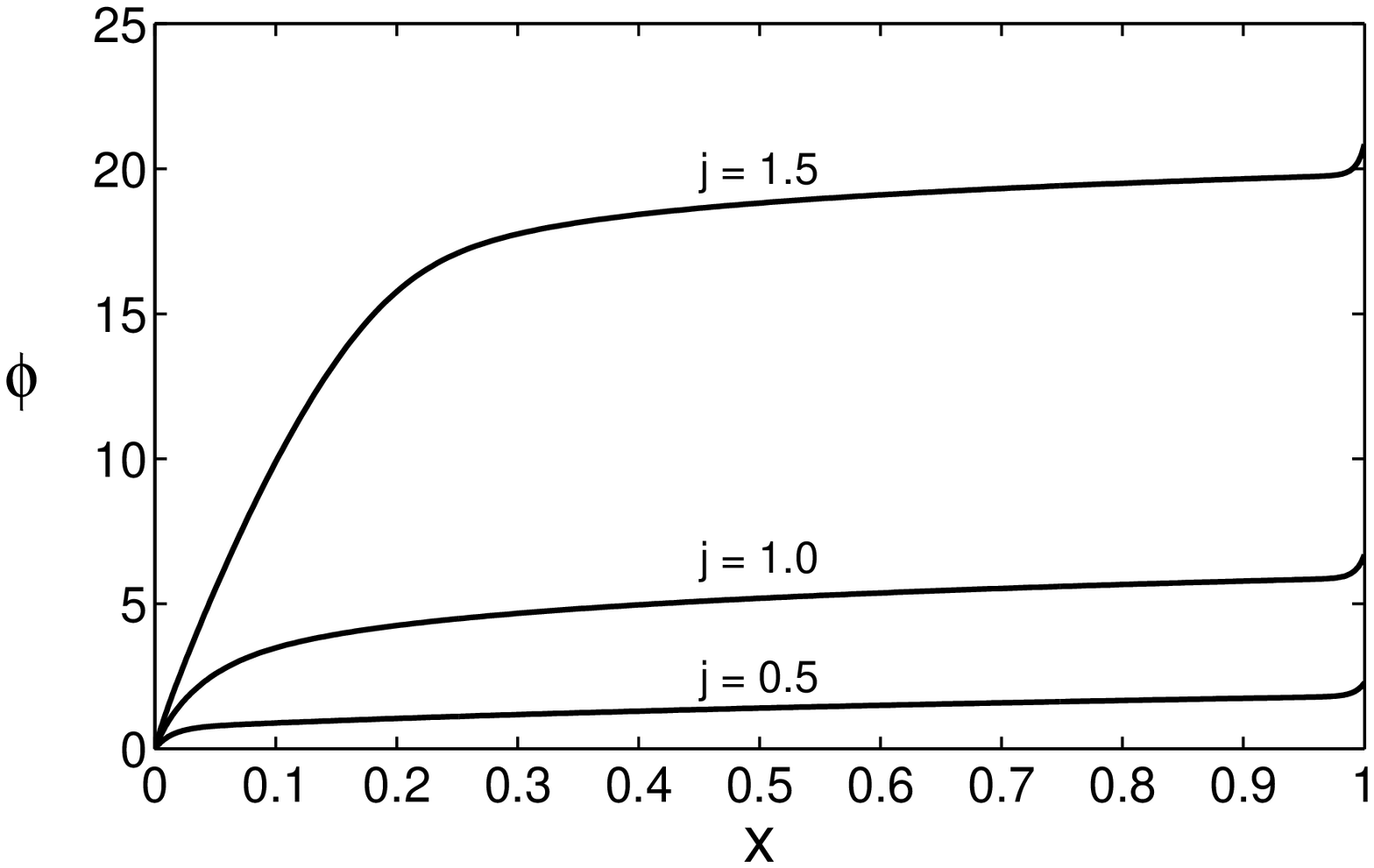}} \nolinebreak
\scalebox{0.35}{\includegraphics{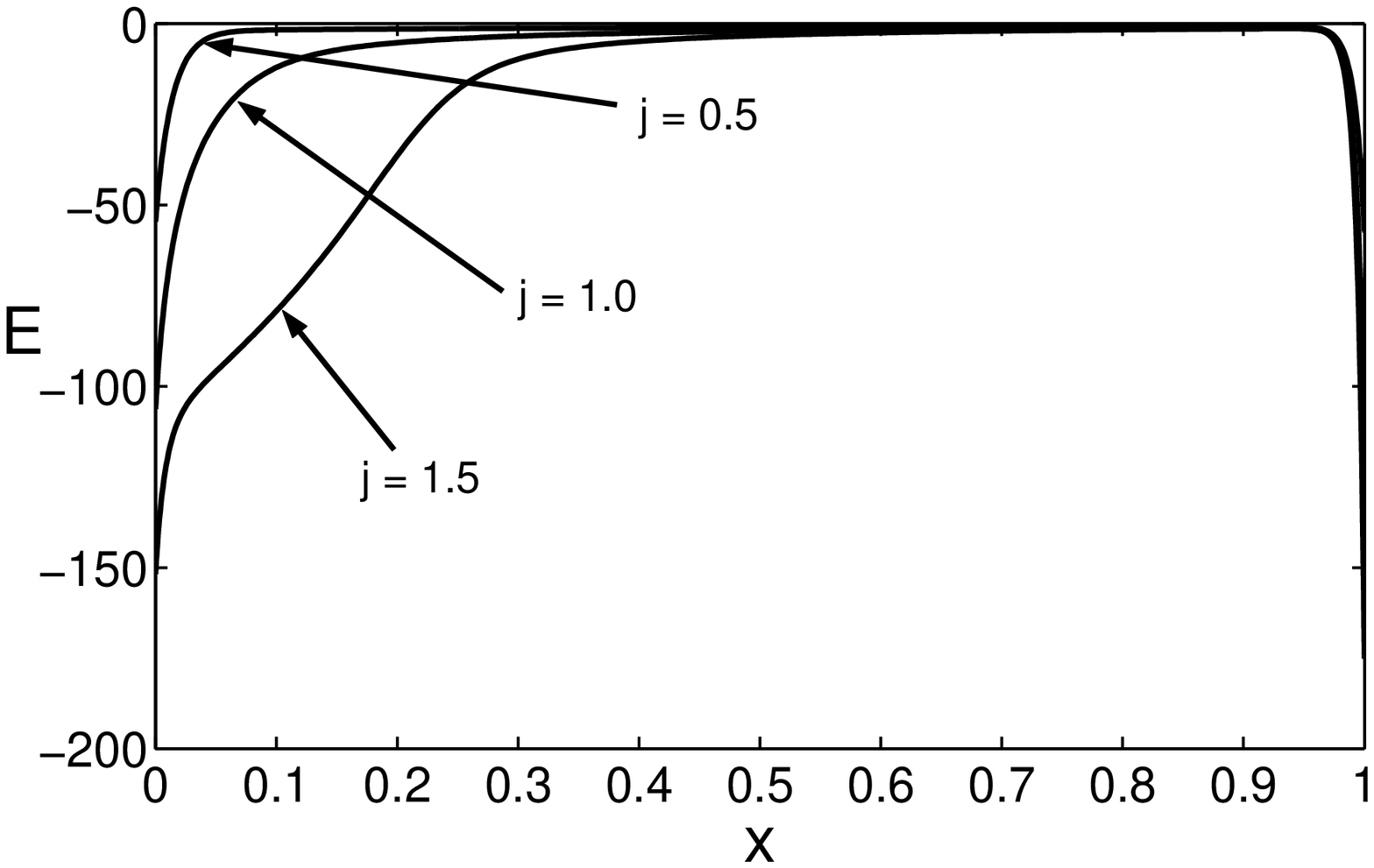}} \\
\scalebox{0.35}{\includegraphics{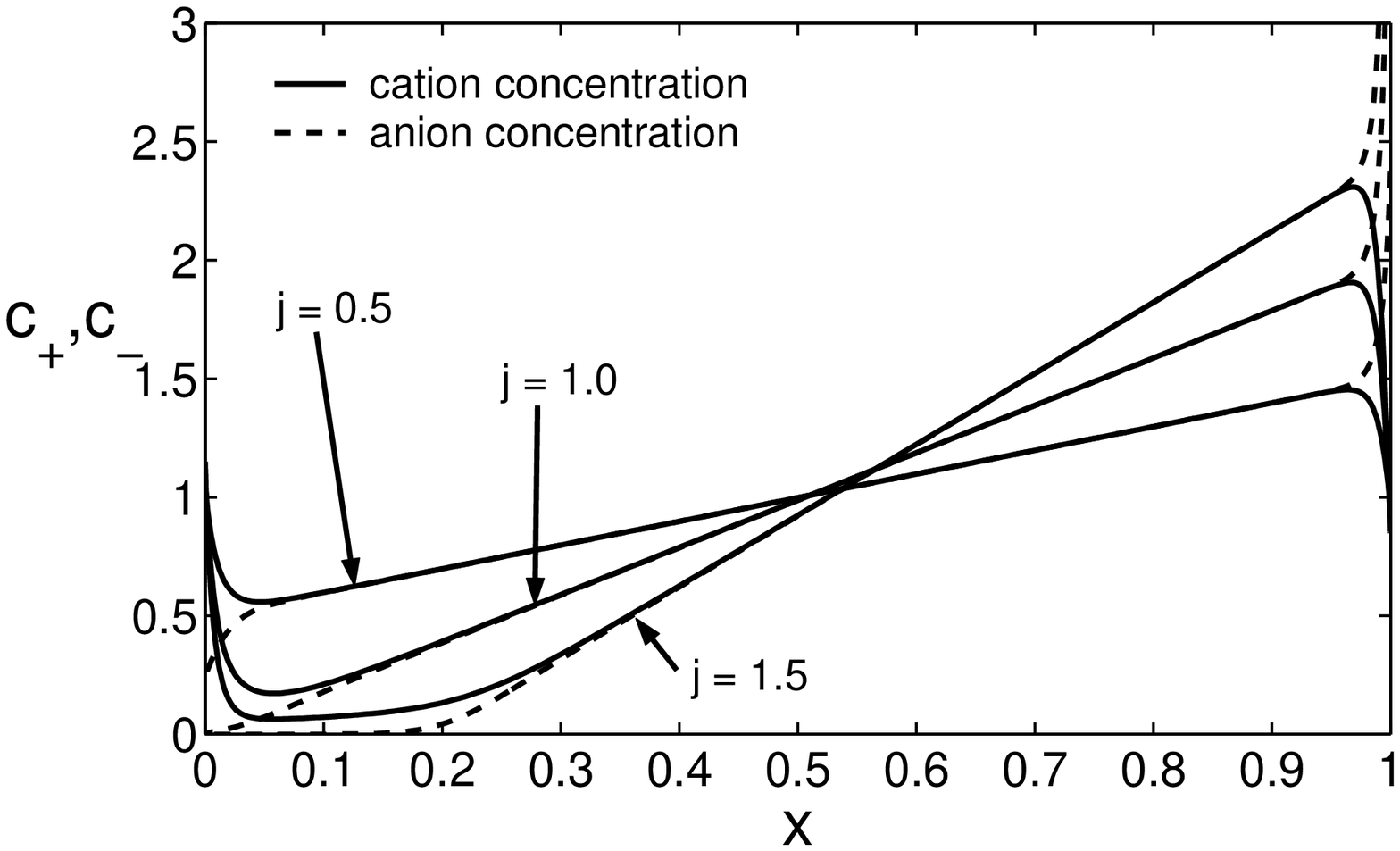}} \nolinebreak
\scalebox{0.35}{\includegraphics{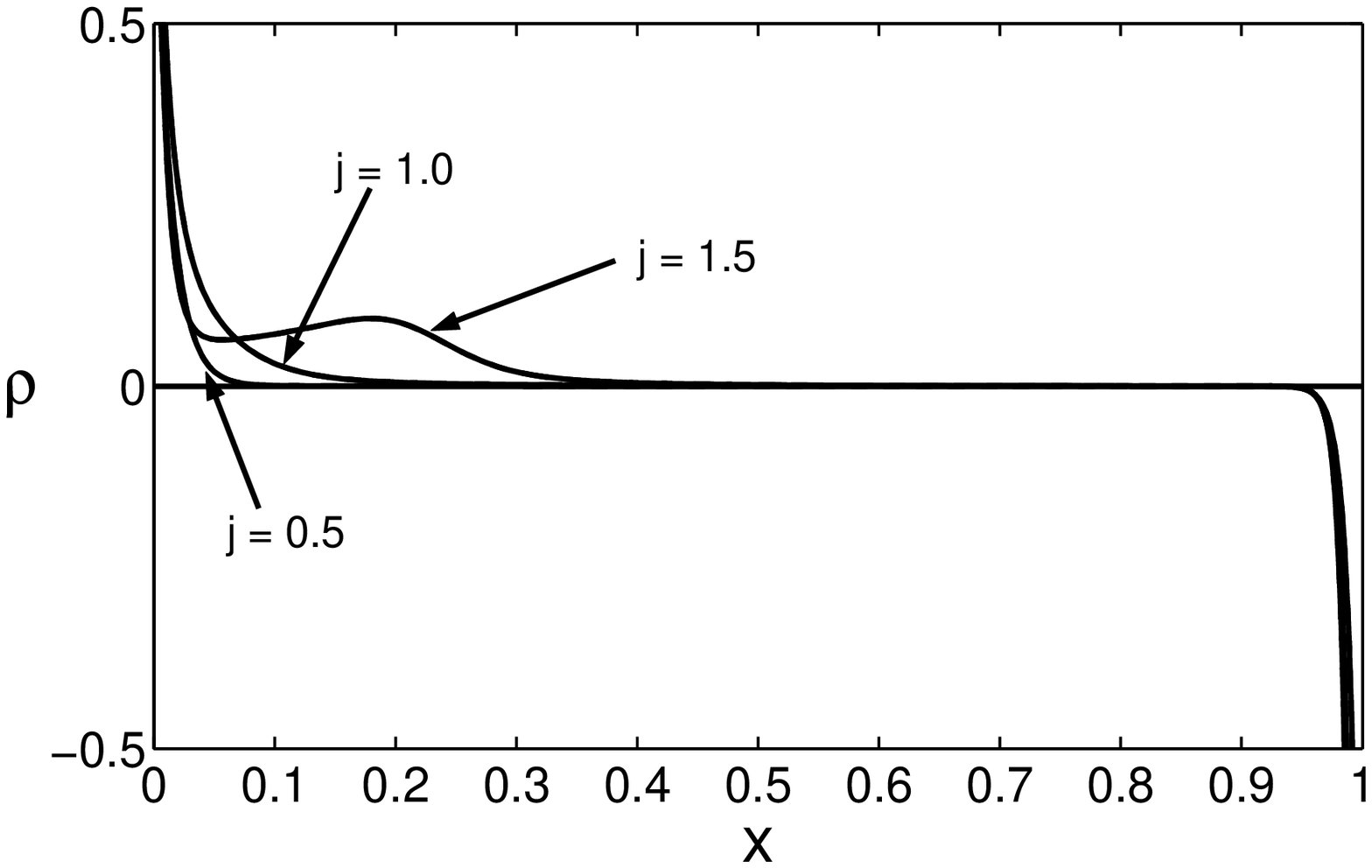}}
\begin{minipage}[h]{5in}
\caption{
\label{figure:compare_fields_at_different_j}
Profiles of the dimensionless potential (top left), electric field
(top right), total ionic concentration (bottom left), and charge
density (bottom right) in three regimes: below the classical
diffusion-limited current ($\i = 0.5$), at the limiting current ($\i =
1$), and above the limiting current ($\i = 1.5$).  These are numerical
solutions to our model problem with the following dimensionelss
parameters: $\epsilon = 0.01$, $\delta = 0$, $k_c = 10$, $\ir = 10$.
}
\end{minipage}
\ec
\end{figure}

The concept of a ``limiting current'', due to the maximum,
steady-state flux of diffusion across an electrochemical cell, was
introduced by Nernst a century ago~\cite{nernst1904}, but
it was eventually realized that the classical theory is flawed, as
illustrated in Figure~\ref{figure:compare_fields_at_different_j} by
numerical solutions to our model problem below. Levich was perhaps the
first to notice that the assumption of bulk electroneutrality yields
approximate solutions to the Poisson-Nernst-Planck (PNP) equations,
which are not self consistent near the limiting current, since the
predicted charge density eventually exceeds the salt concentration
near the cathode~\cite{levich_book}.  This paradox was first resolved
by Smyrl and Newman, who showed that the double layer expands at the
limiting current, as the Poisson-Boltzmann approximation of thermal
equilibrium breaks down~\cite{smyrl1967}.  Rubinstein and Shtilman
later pointed out that mathematical solutions also exist for larger
currents, well above the classical limiting value, characterized by a
region of non-equilibrium ``space charge'' extending significantly
into the neutral bulk~\cite{rubinstein1979}.

The possiblity of superlimiting currents has been studied extensively
in the different context of bulk liquid electrolytes, where a thin
space-charge layer drives nonlinear electro-osmotic slip. This
phenomenon of ``electro-osmosis of the second kind'' was introduced by
Dukhin for the nonlinear electrophoresis of ion-selective, conducting
colloidal particles~\cite{dukhin1991}, and Ben and Chang have recently
studied it in microfluidics~\cite{ben2002}. The mathematical analysis
of second-kind electro-osmosis using matched asymptotic expansions,
similar to the approach taken here, was first developed by Rubinstein
and Zaltzman for related phenomena at electrodialysis
membranes~\cite{rubinstein2000,rubinstein2001}.  In earlier studies,
the space-charge layer was also invoked by Bruinsma and
Alexander~\cite{bruinsma1990} to predict hydrodynamic instability
during electrodeposition and by Chazalviel~\cite{chazalviel1990} in a
controversial theory of fractal electrochemical growth.

As in our companion paper on sublimiting currents~\cite{part1}, here
we consider (typically solid or gel) thin films, e.g. arising in
micro-batteries, which approach the classical limiting current without
hydrodynamic instability. At micron or smaller length scales, the
space charge layer need not be ``thin'' compared to the film
thickness, so we also analyze currents well above the classical
limiting current, apparently for the first time.  In both regimes,
close to and far above the classical limiting current, we derive
matched asymptotic expansions for the concentration profiles and
potential, which we compare against numerical solutions. In addition
to our focus on superlimiting currents and small systems, a notable
difference with the literature on second-kind electro-osmosis is our
use of nonlinear boundary conditions for Faradaic electron-transfer
reactions, assuming Butler-Volmer kinetics and a compact Stern
layer. We also analyze the current-voltage relation, thus extending
our analogous results for thin films below the limiting
current~\cite{part1}.

%

%

\section{Statement of Problem}
Before delving into the analysis (and to make this paper as
self-contained as possible), we review governing equations and
boundary conditions.  We shall focus solely on the dimensionless
formulation of the problem.  For details of the physical assumptions
underlying the mathematical model, the reader is referred
to Ref.~\cite{part1}.

The transport of cations and anions is described by the
steady Nernst-Planck equations 
\bea
\frac{d^2c_+}{dx^2} + \frac{d}{dx}\left( c_+ \frac{d\phi}{dx}
\right) &=& 0 \label{eq:NP_cation} \\
\frac{d^2c_-}{dx^2} - \frac{d}{dx}\left( c_- \frac{d\phi}{dx}
\right) &=& 0.  \label{eq:NP_anion} 
\eea
while Poisson's equation relates the electric potential to the charge density, 
\bea
- \epsilon^2 \frac{d^2\phi}{dx^2} = \frac{1}{2}(c_+ - c_-).  \label{eq:poisson}
\eea
Here $\epsilon$ is a small dimensionless parameter equal to the ratio
of the Debye screening length to the electrode separation (or film
thickness). Note that this formulation assumes constant material
properties, such as diffusivity, mobility, and dielectric coefficient,
and neglects any variations which may occur at large electric fields.
The factor of $1/2$ multiplying the charge density $c_+ - c_-$ is
present merely for convenience.  The domain for the system of
eqns.~(\ref{eq:NP_cation})-(\ref{eq:poisson}) is $0 < x < 1$.

The two Nernst-Planck equations are easily integrated under the physical
constraint that the boundaries are impermeable to anions ({\it i.e.} zero 
flux of anions at $x = 0$) and taking the nondimensional current density at 
the electrodes to be $4\i$: 
\bea
\frac{dc_+}{dx} + c_+ \frac{d\phi}{dx} & = & 4 \i
  \label{eq:c+eq}\\
\frac{dc_-}{dx} - c_-\frac{d\phi}{dx} & = & 0 \label{eq:c-eq}.
\eea
Then by introducing the average ion concentration and (half) the charge 
density
\bea
c = \frac{1}{2}(c_+ + c_-) \ \ \ \textrm{and} \ \ \
\rho = \frac{1}{2}(c_+ - c_-) \label{eq:c_and_rho_def},
\eea
we can derive a more symmetric form for the coupled Poisson-Nernst-Planck 
equations:
\bea
\frac{dc}{dx} + \rho \frac{d\phi}{dx} = 2 \i \label{eq:cphieq} \\
\frac{d\rho}{dx} + c \frac{d\phi}{dx} = 2 \i \label{eq:rhophieq} \\
- \epsilon^2 \frac{d^2\phi}{dx^2} = \rho. \label{eq:phieq}
\eea

For this system of one second-order and two first-order differential 
equations, we require four boundary conditions and one integral constraint:
\bea
\phi(0) - \delta \epsilon \frac{d\phi}{dx}(0) & = & 0 ,
\label{eq:potbc0} \\
\phi(1) + \delta \epsilon \frac{d\phi}{dx}(1) & = & v
\label{eq:potbc1} , \\
k_c[c(0) + \rho(0)]e^{\alpha_c\phi(0)} - \ir e^{-\alpha_a\phi(0)} & = & \i
 \label{eq:potbvbc0} , \\
-k_c[c(1) + \rho(1)]e^{\alpha_c(\phi(1)-v)} + \ir
e^{-\alpha_a(\phi(1)-v)} & = & \i \label{eq:potbvbc1} ,  \\
\int_0^1 [c(x) - \rho(x)]dx & = & 1  ,\label{eq:cint}
\eea
The first two boundary conditions,
Eqs.~(\ref{eq:potbc0})--(\ref{eq:potbc1}), account for the intrinsic
capacitance of the compact part of the electrode-electrolyte
interface, as originally envisioned by Stern. In these boundary
conditions, $\delta$ is a dimensionless parameter which is a measure
of the strength of the surface capacitance, and $v$ is the total
voltage drop across the cell.  

The next two boundary conditions,
Eqs.~(\ref{eq:potbvbc0})--(\ref{eq:potbvbc1}), are Butler-Volmer rate
equations which represent the kinetics of Faradaic electron-transfer
reactions at each electrode, with an Arhhenius dependence on the
compact layer voltage.  In these equations, $k_c$ and $\ir$ are
dimensionless reaction rate constants and $\alpha_c$ and $\alpha_a$
are transfer coefficients for the electrode reaction.  It is worth
noting that $\alpha_c$ and $\alpha_a$ do not vary too much from system
to system; typically they have values between $0$ and $1$ and often
both take on values near $1/2$.

Finally, the integral constraint, Eq.~(\ref{eq:cint}), reflects the
fact that the total number of anions is fixed, assuming that anions
are not allowed to leave the electrolyte by Faradaic processes or
specific adsorption.  It is important to understand that the need for
an {\it extra} boundary condition/constraint reflects that the
current-voltage relationship, $\i(v)$, or ``polarographic curve'', is
not given {\it a priori}.  As usual in one-dimensional
problems~\cite{part1}, it is easier to assume galvanostatic forcing at
fixed curent, $\i$, and then solve for the cell voltage, $v(\i)$, by
applying the boundary condition (\ref{eq:potbc1}), rather than the
more common case of potentiostatic forcing at fixed voltage, $v$.
For this reason, we take the former approach in our analysis.
For steady-state
problems, the two kinds of forcing are equivalent and yield the same
(invertible) polaragraphic curve, $\i(v)$ or $v(\i)$.

For some of our analysis, it will be convenient to further
simplify the problem by introducing the dimensionless electric field,
$E \equiv -\frac{d\phi}{dx}$.
This transformation is useful because three of the five independent
constraints can be expressed in terms of these variables,
without explicit dependence on
$\phi(x)$, namely the two Butler-Volmer rate equations,
\bea
k_c (c(0) + \rho(0)) e^{-\alpha_c\delta\epsilon E(0)}
        - \ir e^{\alpha_a\delta\epsilon E(0)} & = & \i
        \label{eq:bv0} , \\
- k_c (c(1) + \rho(1)) e^{\alpha_c\delta\epsilon E(1)}
        + \ir e^{-\alpha_a\delta\epsilon E(1)} & = & \i
        \label{eq:bv1} ,
\eea
and the integral constaint on the total number of anions,
Eq.~(\ref{eq:cint}).
The potential is recovered by integrating the electric field and applying
the Stern boundary conditions Eqs.~(\ref{eq:potbc0}) and (\ref{eq:potbc1}).

\section{Unified Analysis at All Currents}

\subsection{Master Equation for the Electrostatic Potential}
We begin our analysis by reducing the governing equations,
Eqs.~(\ref{eq:cphieq}) through (\ref{eq:phieq}), to
a single master equation for the electrostatic potential.
Substituting Eq.~(\ref{eq:phieq}) into Eq.~(\ref{eq:cphieq}) and integrating,
we obtain an expression for the average concentration
\beq
c(x) =   \b{c}_o + 2 \i x + \frac{\epsilon^2}{2} \left ( \frac{d\phi}{dx} \right )^2 .
  \label{eq:c_master_eqn}
\eeq
Then by applying the integral constraint, Eq.~(\ref{eq:cint}), we find that
the integration constant, $\b{c}_o$ , is given by 
\beq
\b{c}_o = (1-\i)
  - \epsilon^2 \left [
  \left ( \left. \frac{d\phi}{dx} \right ) \right|_{x=1} 
  - \left ( \left. \frac{d\phi}{dx} \right ) \right|_{x=0} 
  +  \frac{1}{2}\int_0^1 \left ( \frac{d \phi}{dx} \right)^2 dx
  \right ]
  \label{eq:c_o_master}
\eeq
Note that when the electric field is $O(1)$, Eqns.~(\ref{eq:c_master_eqn}) 
and (\ref{eq:c_o_master}) reduce to the leading-order concentration 
in the bulk when $\i$ is sufficiently below the limiting 
current~\cite{part1}.
We can now eliminate $\rho$ and $c$ from Eq.~(\ref{eq:rhophieq}) to 
arrive at a single master equation for $\phi$
\beq
  \epsilon^2 
  \left [ 
    -\frac{d^3 \phi}{dx^3} 
    + \frac{1}{2} \left ( \frac{d \phi}{dx} \right )^3
  \right ]
  + \left (\b{c}_o + 2 \i x \right ) \frac{d \phi}{dx} = 2 \i,
  \label{eq:phi_master_eqn}
\eeq
or equivalently for the electric field $E$
\beq
  \epsilon^2 
  \left [ 
    \frac{d^2 E}{dx^2} 
    - \frac{1}{2} E^3
  \right ]
  - \left (\b{c}_o + 2 \i x \right ) E = 2 \i.
  \label{eq:E_master_eqn}
\eeq
Once this equation is solved, the concentration, $c$, and 
charge density, $\rho$, are computed using Eq.~(\ref{eq:c_master_eqn}) and
Poisson's equation, Eq.~(\ref{eq:phieq}).   

The master equation has been derived in various equilivalent forms
since the 1960s. Grafov and Chernenko~\cite{grafov1962} first combined
Eqs.~(\ref{eq:c+eq}), (\ref{eq:c-eq}) and (\ref{eq:phieq}) to obtain a
single nonlinear differential equation for the anion concentration,
$c_-$, whose general solution they expressed in terms of Painlev\'e's
transcendents. The master equation for the electric field,
Eq.~(\ref{eq:E_master_eqn}), was first derived Smyrl and
Newman~\cite{smyrl1967}, in the special case of the classical limiting
current, where $\i=1$ and $\b{c}_o=0$, where they discovered a
non-equilibrium double layer of width, $\epsilon^{2/3}$, which is
apparent from the form of the master equation. We shall study the
general electric-field and potential equations for an arbitrary
current, $\i$, focusing on boundary-layer structure in the limiting
and superlimiting regimes.

\subsection{ Efficient Numerical Solution}
\label{app:num_methods}
To solve the master equation for the electric field with the boundary
conditions and integral constraint, we use the Newton-Kantorovich
method~\cite{boyd_book}. Specifically, we use a Chebyshev
pseudospectral discretization to solve the linearized boundary-value
problem at each iteration~\cite{boyd_book,trefethen_book}.  Our
decision to use this method is motivated by its natural ability to resolve
boundary layers and its efficient use of grid points.  We are able to
get accurate results for many parameter regimes very quickly
(typically less than a few minutes on a workstation) with only a few
hundred grid points, which would not be possible at large currents
and/or thin double layers using a naive finite-difference scheme.
It is important to stress that the boundary conditions and the
integral constraint are explicitly included as part of the
Newton-Kantorovich iteration.  Therefore, the linear BVP solved in
each iteration is actually an {\it integro-differential
differential equation} with boundary conditions that are
integro-algebraic equations.

To ensure convergence at high currents, we use continuation in the
current density parameter, $\i$, and start with a sufficiently low
initial $\i$ that the bulk electroneutral solution is an acceptable
initial guess; often, initial $\i$ values relatively high compared to
the diffusion-limited current are acceptable.  Continuation in the
$\delta$ parameter is also sometimes necessary to compute solutions at
high $\delta$ values.

The results of the numerical method are presented in the figures
below and in Ref.~\cite{part1} to test our analytical
approximations obtained by asymptotic analysis.

\subsection{Recovery of Classical Results Below the Limiting Current, $\i \ll 1 - \e23$} 
In the low-current regime, the master equation admits the two
distinguished limits around $x=0$ that arise in the classical
analysis: $x = O(1)$ and $x = O(\epsilon)$.  When $x=O(1)$, we find
the usual bulk electric field from Eq.~(\ref{eq:phi_master_eqn})
and the bulk concentration from Eq.~(\ref{eq:c_master_eqn}).  When
$x=O(\epsilon)$, the master equation can be rescaled using $x=\epsilon
y$ to obtain
\beq
  -\frac{d^3 \phi}{dy^3} + \frac{1}{2} \left (\frac{d \phi}{dy} \right )^3
  + \b{c}_o \frac{d \phi}{dy} + 2 \i y \epsilon \frac{d \phi}{dy} 
  = 2 \i \epsilon 
\eeq
which is equivalent to the classical theory at leading
order~\cite{part1}. In particular, the Gouy-Chapman structure of the
double-layer can be derived directly from the Smyrl-Newman equation in
this limit~\cite{bonnefont2001}.

The anode boundary layer comes from a similar $O(\epsilon)$ scaling
around $x=1$.  Note that in the $\i \ll 1-\e23$ regime, the scaling
$x=O(\e23)$ is {\it not} a distinguished limit because the
$\b{c}_o \left ( \frac{d \phi}{dx} \right )$ term would dominate all 
other terms in Eq.~(\ref{eq:phi_master_eqn}).

\section{Nested Boundary Layers at the Limiting Current, $\i = 1 -
O(\e23)$}  In this section, we show that a nontrivial nested
boundary-layer structure emerges at the classical limiting current
when general boundary conditions are considered.

\subsection{ Expansion of the Double Layer Out of Equilibrium}
\begin{figure}
\bc
\scalebox{0.5}{\includegraphics{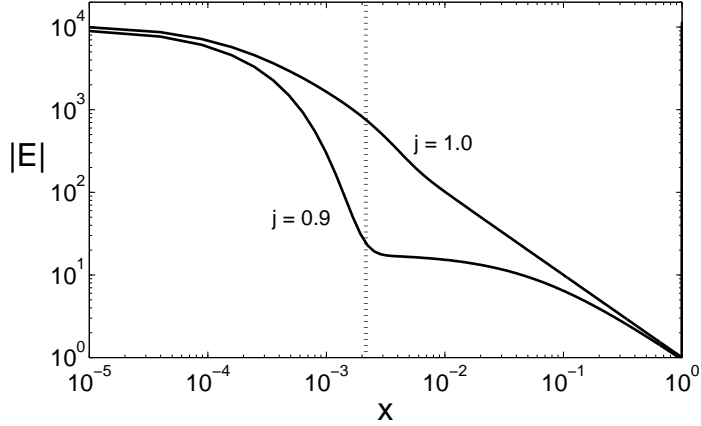}}
\begin{minipage}[h]{5in}
\caption{
\label{figure:diffuse_layer_expansion}
Numerical solutions for the dimensionless electric field $E(x)$ at
current densities of $\i=0.9$ and $\i=1.0$ demonstrating the expansion
of the diffuse layer at the limiting current ($k_c = 1$, $\ir = 2$,
$\delta=0.1$ and $\epsilon = 0.0001$).  For reference, the vertical
line shows where $x = \epsilon^{2/3}$. }
\end{minipage}
\ec
\end{figure}

As discussed in the companion paper~\cite{part1}, the classical
analysis breaks down as the current approaches the diffusion-limited
current.  Mathematically, the breakdown occurs because a new
distinguished limit for the master equation appears as $\i \rightarrow
1$.  Rescaling the master equation using $x=\e23 z$ gives us
\beq
  -\frac{d^3 \phi}{dz^3} + \frac{1}{2} \left (\frac{d \phi}{dz} \right )^3
  + \frac{\b{c}_o}{\e23} \frac{d \phi}{dz} 
  + 2 \i z \frac{d \phi}{dz} 
  = 2 \i,
  \label{eq:e23_master_eqn}
\eeq
which implies that we have a meaningful distinguished limit if
$\b{c}_o = O(\e23)$ or, equivalently, $\i = 1-O(\e23)$.  As first
noticed by Smyrl and Newman~\cite{smyrl1967}, the appearance of the
$\e23$ scaling corresponds to the expansion of the diffuse charge
layer that arises at the classical limiting current (see
Figure~\ref{figure:diffuse_layer_expansion}). In this regime, the
double layer is no longer in Poisson-Boltzmann equilibrium at
leading order.

Unfortunately, at this scale, {\it all} terms in
Eq.~(\ref{eq:e23_master_eqn}) are $O(1)$, so we are forced to solve
the full equation. Although general solutions can be expressed in
terms of Painlev\'e's transcendents~\cite{rubinstein_book,ben2002,grafov1962},
these are not convenient for applying our nonlinear boundary
conditions or obtaining physical insight.  Even when $\b{c}_o =
o(\e23)$, we are left with a complicated differential equation which
does not admit a simple analytical solution.  However, in the case
$\b{c}_o = o(\e23)$, it is possible to study the asymptotic behavior
of the solution in the limits $z \rightarrow 0$ and $z \rightarrow
\infty$ by considering the behavior of the {\it neighboring}
asymptotic layers.

\subsection{Nested Boundary Layers when $|1-\i| = o(\e23)$} 
The appearance of the new distinguished limit for $\i = 1-O(\e23)$
does not destroy the ones that exist in the classical analysis.  In
particular, the $O(\epsilon)$ boundary layer at $x=0$ does {\it not}
vanish. This inner layer was overlooked by Smyrl and Newman because
they assumed a fixed surface charge density given by the equilibrium
zeta potential~\cite{smyrl1967}, rather than more realistic boundary
conditions allowing for surface-charge variations. 

\begin{figure}
\bc
\scalebox{0.5}{\includegraphics{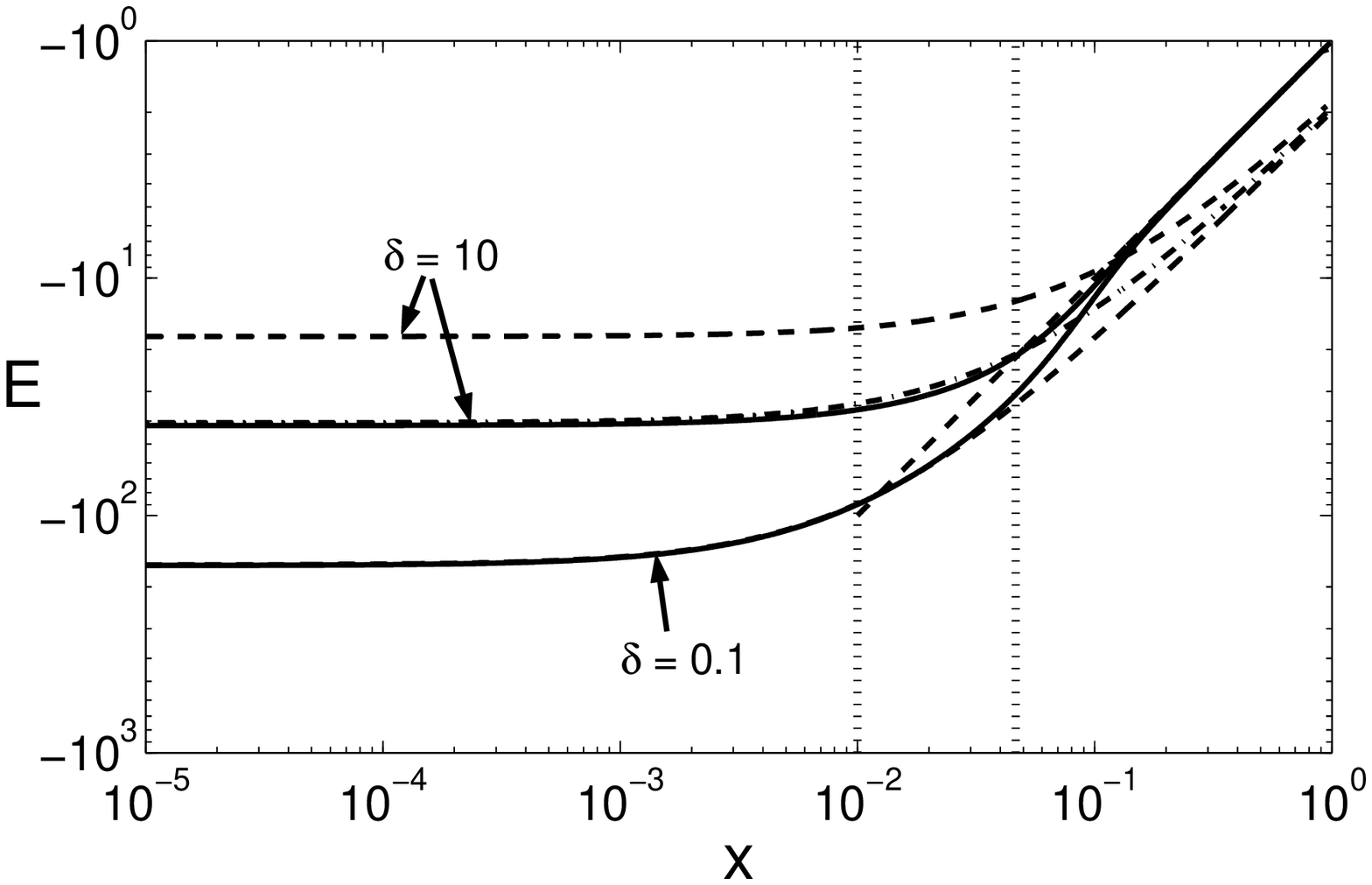}} \\
\scalebox{0.5}{\includegraphics{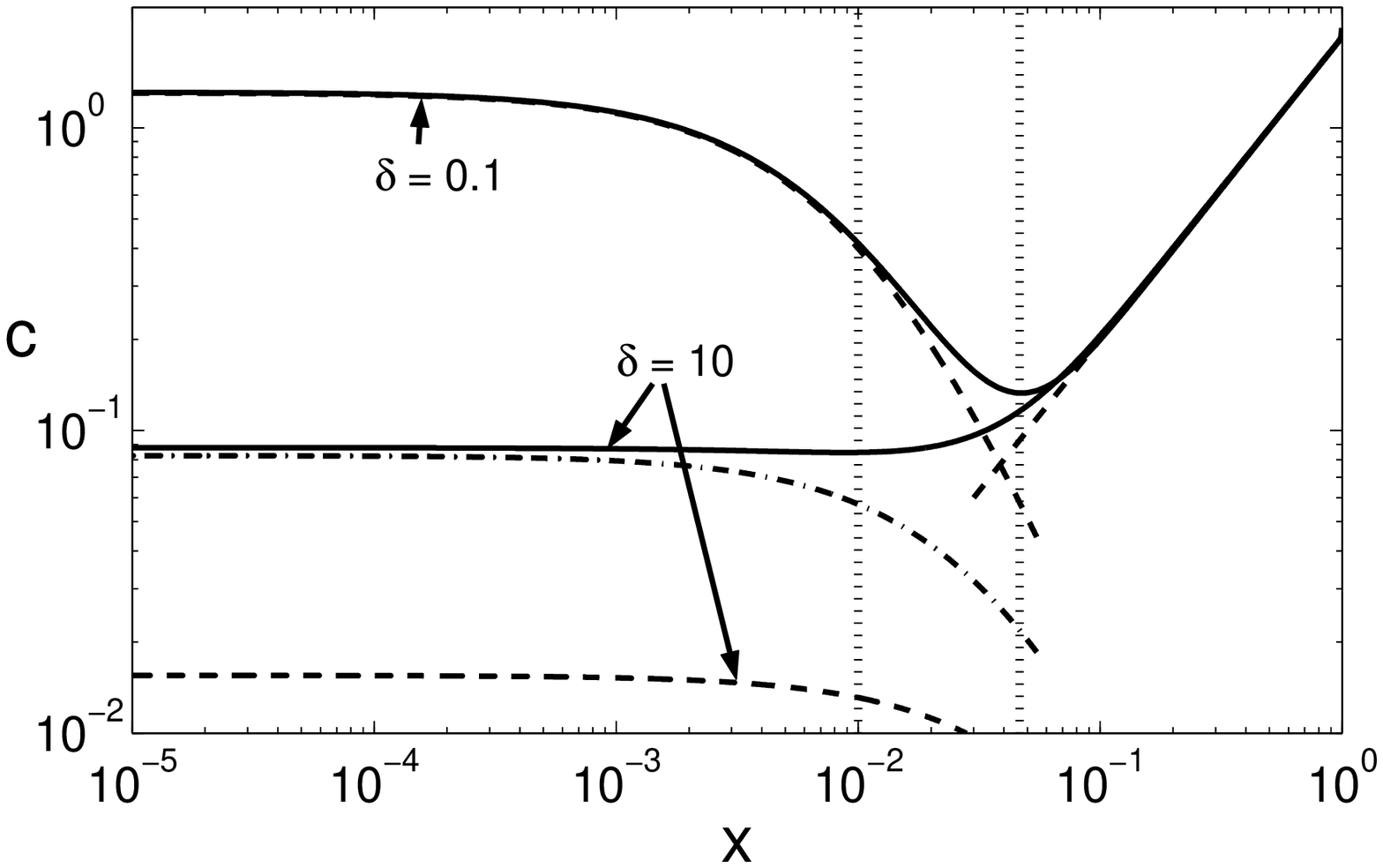}}
\begin{minipage}[h]{5in}
\caption{
\label{figure:fields_j1_0_e0_01}
Numerical solutions (solid lines) for the dimensionless electric field
$E(x)$ and concentration $c(x)$ at the classical diffusion-limited
current ($\i = 1$) compared with leading order asymptotic
approximations (dashed and dot-dashed lines) for $k_c = 1$, $\ir = 2$,
$\epsilon = 0.01$ and $\delta = 0.1 , 10$.  The leading order bulk
approximations for $E(x)$ and $c(x)$ are given by
Eq.~(\ref{eq:bulk_asym_sol_SN}) and $c(x) = 2\i x$, respectively.  In
the diffuse layer, the leading order approximations are given by
Eqs.~(\ref{eq:E_inner_diffuse}) and (\ref{eq:C_inner_diffuse}).  For
the $\delta = 10$ curves, the difference between the dashed and
dot-dashed curves is that the dahsed curve uses an approximate value
for $B$ given by Eq.~(\ref{eq:B_large_delta}) while the dot-dashed
curve uses a $B$ value calculated by numerically solving
Eq.~(\ref{eq:b_eqn}).  For reference, the vertical lines show where $x
= \epsilon$ and $x = \e23$.  The thin anode diffuse layer field is not
shown.}
\end{minipage}
\ec
\end{figure}

In the general case, a set of nested boundary layers exists when the
current is near the classical limiting current.  For convenience, we
shall refer to the $x = O(\e23)$ and the $x = O(\epsilon)$ regions as
the ``Smyrl-Newman'' and ``inner diffuse'' layers, respectively.  It
is important to realize that without the $O(\epsilon)$ layer, it would
be impossible, in general, to satisfy the boundary conditions of the
original problem.  To see this, consider the reaction boundary
condition at $x=0$, Eq.~(\ref{eq:potbvbc0}).  To estimate the $c$ and
$\rho$ at the electrode surface, we rescale
Eq.~(\ref{eq:c_master_eqn}) and Poisson's equation using $x=\e23 z$ to
obtain
\bea
  c & = & \b{c}_o + 2 \i \e23 z 
   + \frac{\e23}{2} \left (\frac{d\phi}{dz}\right )^2 \\
  \rho & = & -\e23 \frac{d^2 \phi}{d z^2},
\eea
which means that the concentration and charge density are both 
$O(\e23)$ since $\b{c}_o = o(\e23)$ when $|1-\i| = o(\e23)$.  
Then, from the Stern boundary condition, we have 
$\phi(0) = -\delta \epsilon \Eb = 
-\delta \epsilon^{1/3} \Eg = O(\delta \epsilon^{1/3})$.
Plugging these estimates into the reaction boundary condition, we find 
\beq
  k_c O(\e23) e^{\alpha_c \delta \epsilon^{1/3} \Eg(0)} = \i + 
    \ir e^{-\alpha_a \delta \epsilon^{1/3} \Eg(0)} = O(1).
\eeq
For this equation to be satisfied, $\delta$ is constrained to be
huge for $\epsilon$ near $0$:
$\delta = O\left ( \left | \log \e23 \right | /\epsilon^{1/3} \right ) \gg 1$.
But $\delta$ is an independently assigned physical parameter, 
so there is no reason that it should have to satisfy this constraint.
Thus, we are lead to the conclusion that there {\it must} be a 
nested inner boundary layer in order to satisfy the Butler-Volmer 
boundary condition when $\delta$ is small compared to 
$\left | \log \e23 \right | / \epsilon^{1/3}$.
In other words, for insufficiently large $\delta$,
the reaction boundary condition requires that the cation concentration 
is $O(1)$ at $x=0$, which implies the existence of a boundary layer between 
the Smyrl-Newman layer and the boundary.  
As can be seen in figure \ref{figure:fields_j1_0_e0_01},
the cathode layer concentration decreases significantly as $\delta$ is 
increased.

To analyze Eq.~(\ref{eq:e23_master_eqn}), it is convenient to focus on 
the electric field rather than the potential.  
In terms of the scaled electric field, 
$\Eg(z) \equiv -\frac{d\phi}{dz} = \e23 E(x)$, 
Eq.~(\ref{eq:e23_master_eqn}) becomes
\beq
  \frac{d^2 \Eg}{dz^2} - \frac{1}{2} \Eg^3
  - 2 \i \left ( z \Eg + 1 \right )
  = \frac{\b{c}_o}{\e23} \Eg
  \label{eq:SN_eqn}
\eeq
which we shall refer to as the ``Smyrl-Newman equation''.  
From Eq.~(71) in Ref.~\cite{part1}, we know that the
first few terms in the expansion for the bulk electric field at the 
limiting current are 
\bea
  -\Eb(x) &=& \frac{1}{x}
            + \frac{3 \epsilon^2}{4 x^4} 
            + \frac{111 \epsilon^4}{16 x^7} 
            + \frac{6045 \epsilon^6}{32 x^{10}} 
            + \ldots \nonumber \\
         &=& \frac{1}{\e23} \left (
	      \frac{1}{z}
            + \frac{3}{4 z^4} 
            + \frac{111}{16 z^7} 
            + \frac{6045}{32 z^{10}} 
            + \ldots \right ) .
  \label{eq:bulk_asym_sol_SN}
\eea 
Since the second series is asymptotic for $z \gg 1$, the expansion 
in the bulk is valid for $x \gg \e23$.  In order to match the
solution in the Smyrl-Newman layer to the bulk, we expect 
the asymptotic solution to Eq.~(\ref{eq:SN_eqn}) as 
$z \rightarrow \infty$ to be given by the expression in parentheses 
in Eq.~(\ref{eq:bulk_asym_sol_SN}).  We could also have arrived 
at this result by directly substituting an asymptotic expansion in $1/z$
and matching coefficients. As we can see in 
Figure~\ref{figure:fields_j1_0_e0_01} 
the leading order term in Eq.(\ref{eq:bulk_asym_sol_SN}) 
is a good approximation to the exact solution in the bulk and is matched
by the solution in the Smyrl-Newman layer as it extends into the bulk.

We now turn our attention towards the ``inner diffuse'' layer which gives us 
the asymptotic behavior of the Smyrl-Newman equation in the limit 
$z \rightarrow 0$.  Introducing the scaled variables 
$y = x/\epsilon = z/\epsilon^{1/3}$ and 
$\Ebr = \epsilon \Eb = \epsilon^{1/3} \Eg$, Eq.~(\ref{eq:SN_eqn}) becomes
\beq
  \frac{d^2 \Ebr}{dy^2} - \frac{1}{2}\Ebr^3 
  - 2\i \epsilon \left ( y \Ebr + 1 \right ) = \b{c}_o \Ebr.
\eeq
Near the limiting current ({\it i.e.} $\b{c}_o = O(\e23)$), 
$\Ebr$ satisfies 
$\frac{d^2 \Ebr}{dy^2} = \frac{1}{2}\Ebr^3$
at leading order with the boundary condition $\Ebr \rightarrow 0$ 
as $y \rightarrow \infty$ from the matching condition that $\Eg$ 
remains bounded as $z \rightarrow 0$.  
Integrating this equation twice with the observation that 
$\frac{d \Ebr}{d y} > 0$ gives us
\beq
  \Ebr(y) \sim -\frac{2}{y+b}
  \label{eq:E_inner_diffuse}
\eeq
where $b$ is a constant determined by applying the Butler-Volmer reaction 
boundary condition at the cathode.  
We can estimate $\cbr(y)$ and $\rhobr(y)$ by substituing 
Eq.~(\ref{eq:E_inner_diffuse}) into 
Eq.~(\ref{eq:c_master_eqn}) and Poisson's equation to find
\bea
  \cbr(y) &=& \b{c}_o + 2\i x + \frac{\epsilon^2}{2} \Eb(x)^2 
    = \b{c}_o + 2\i \epsilon y + \frac{1}{2} \Ebr(y)^2 
    = \frac{2}{\left ( y+b \right )^2} + O(\epsilon) 
  \label{eq:C_inner_diffuse} \\
  \rhobr(y) &=& \epsilon^2 \frac{d \Eb}{dx} 
    = \frac{d \Ebr}{dy} 
    = \frac{2}{\left ( y+b \right )^2} + O(\epsilon).
  \label{eq:rho_inner_diffuse}
\eea 
Therefore, $b$ satisfies the following transcendental equation at leading 
order:
\beq
  k_c \frac{4}{b^2} e^{2 \alpha_c \delta/b} 
  = \i + \ir e^{-2 \alpha_a \delta/b}.
  \label{eq:b_eqn}
\eeq
While this equation does not admit a simple closed form solution, we can 
compute approximate solutions in the limits of small and large $\delta$
values.  
In the small $\delta$ limit, we can linearize Eq.(\ref{eq:b_eqn})
and expand $b$ in a power series in $\delta$ to obtain
\beq
  b \sim 2 \sqrt{\frac{k_c}{\i+\ir}} 
    + \delta \left ( \alpha_c + \frac{\alpha_a \ir}{\i + \ir} \right )
    + O(\delta^2).
  \label{eq:b_small_delta}
\eeq
At the other extreme, for $\delta \gg 1$, 
Eq.~(\ref{eq:b_eqn}) can be approximated by 
\beq
  k_c \frac{4}{b^2} e^{2 \alpha_c \delta/b} \approx \i.
\eeq
Then, using fixed-point iteration on the approximate equation, we 
find that
\beq
  b \sim \frac{2 \alpha_c \delta}
    {\log \kappa - 2 \log \log \kappa 
    + O\left(\log\log\log \delta^2 \right )}
  \label{eq:B_large_delta}
\eeq
where $\kappa \equiv \i \alpha_c^2 \delta^2 / k_c$.
Figure~\ref{figure:fields_j1_0_e0_01} shows that the leading order approximation
Eq.~(\ref{eq:E_inner_diffuse}) is very good in the inner diffuse 
layer as long as an accurate estimate for $b$ is used.  
While the small $\delta$ approximation for $b$ is amazingly good 
(the asymptotic and numerical solutions are nearly indistinguishable),
the large $\delta$ estimate for $b$ is not as good but is only off by an 
$O(1)$ multiplicative factor.  

Before moving on, it is worth noting that the asymptotic behavior
of the concentration and charge density in the Smyrl-Newman layer
as $z \rightarrow 0$ and $z \rightarrow \infty$ suggest that the 
charge density is low throughout the entire Smyrl-Newman layer.  
Figure~\ref{figure:fields_j1_0_e0_01} shows how the Smyrl-Newman layer acts as 
a transition layer allowing the bulk concentration to become small near 
the cathode while still ensuring a sufficiently high cation concentration 
at the cathode surface to satisfy the reaction boundary conditions.
The transition nature of the Smyrl-Newman layer becomes even more 
pronounced for smaller values of $\epsilon$.

\section{Bulk Space Charge Above the Limiting Current, $1 + O(\e23) \ll \i \ll O(1/\epsilon)$} 
As current exceeds the classical limiting value, the overlap region
between the inner diffuse and Smyrl-Newman layers grows to become a
layer having $O(1)$ width.  Following other
authors~\cite{rubinstein1979,chazalviel1990}, we shall refer to this
new layer as the ``{\it space-charge}'' layer because, as we shall
see, it has a non-negligible charge density compared to the rest of
the bulk.  Therefore, in this current regime, the central region of
the electrochemical cell is split into two pieces having $O(1)$ width
separated by a $o(1)$ transition layer.

\begin{figure}
\bc
\scalebox{0.42}{\includegraphics{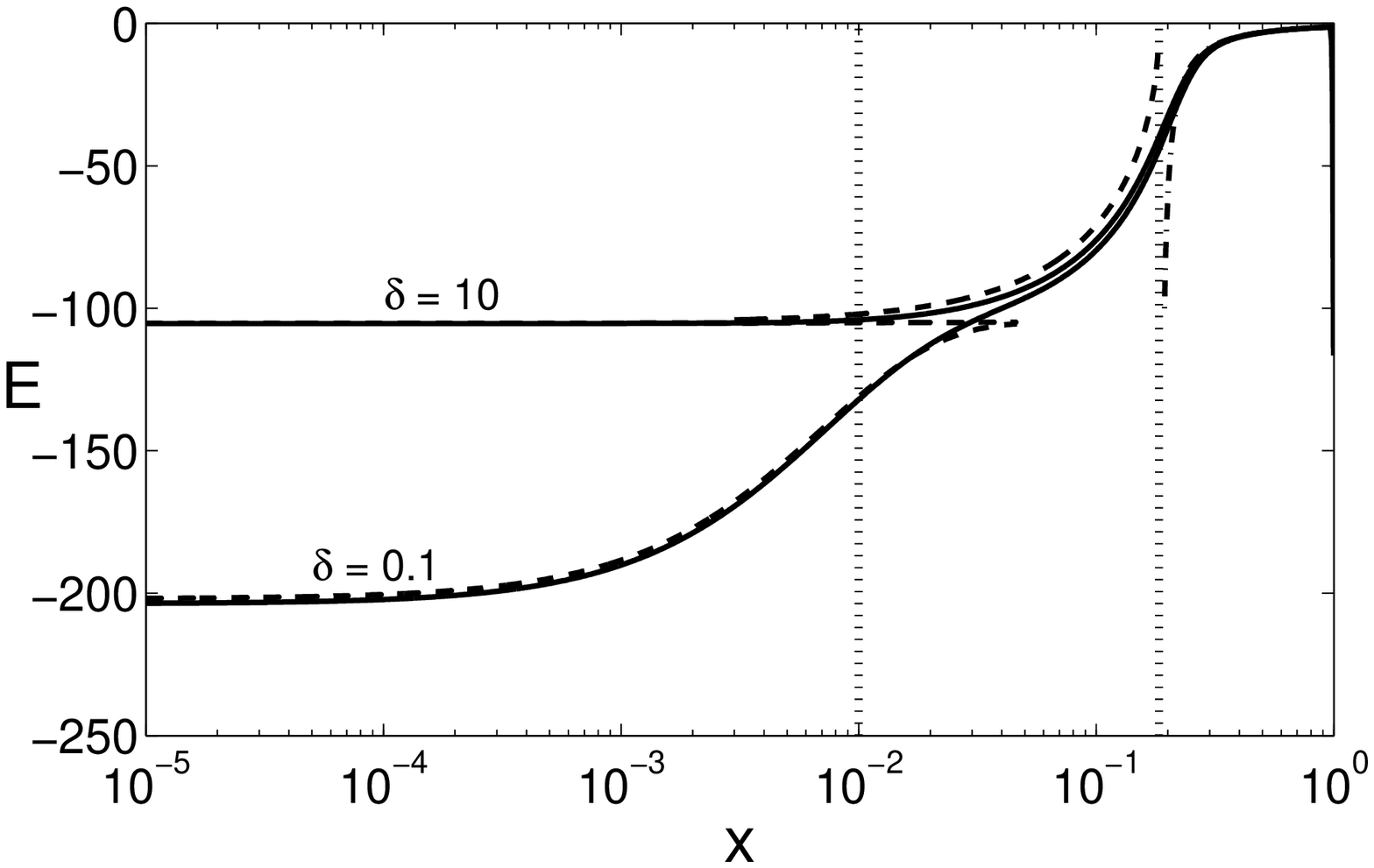}} \\
\scalebox{0.42}{\includegraphics{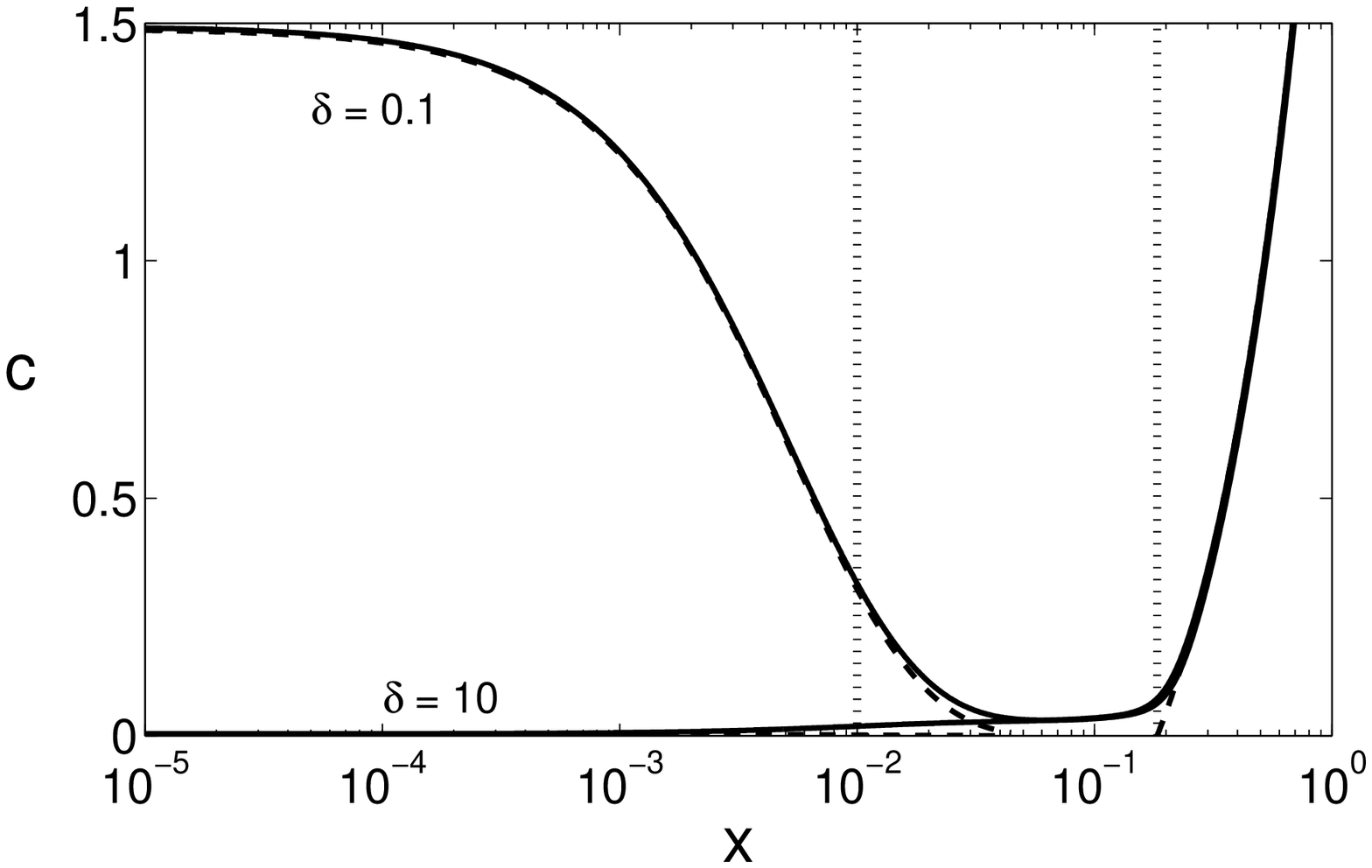}} \\
\scalebox{0.42}{\includegraphics{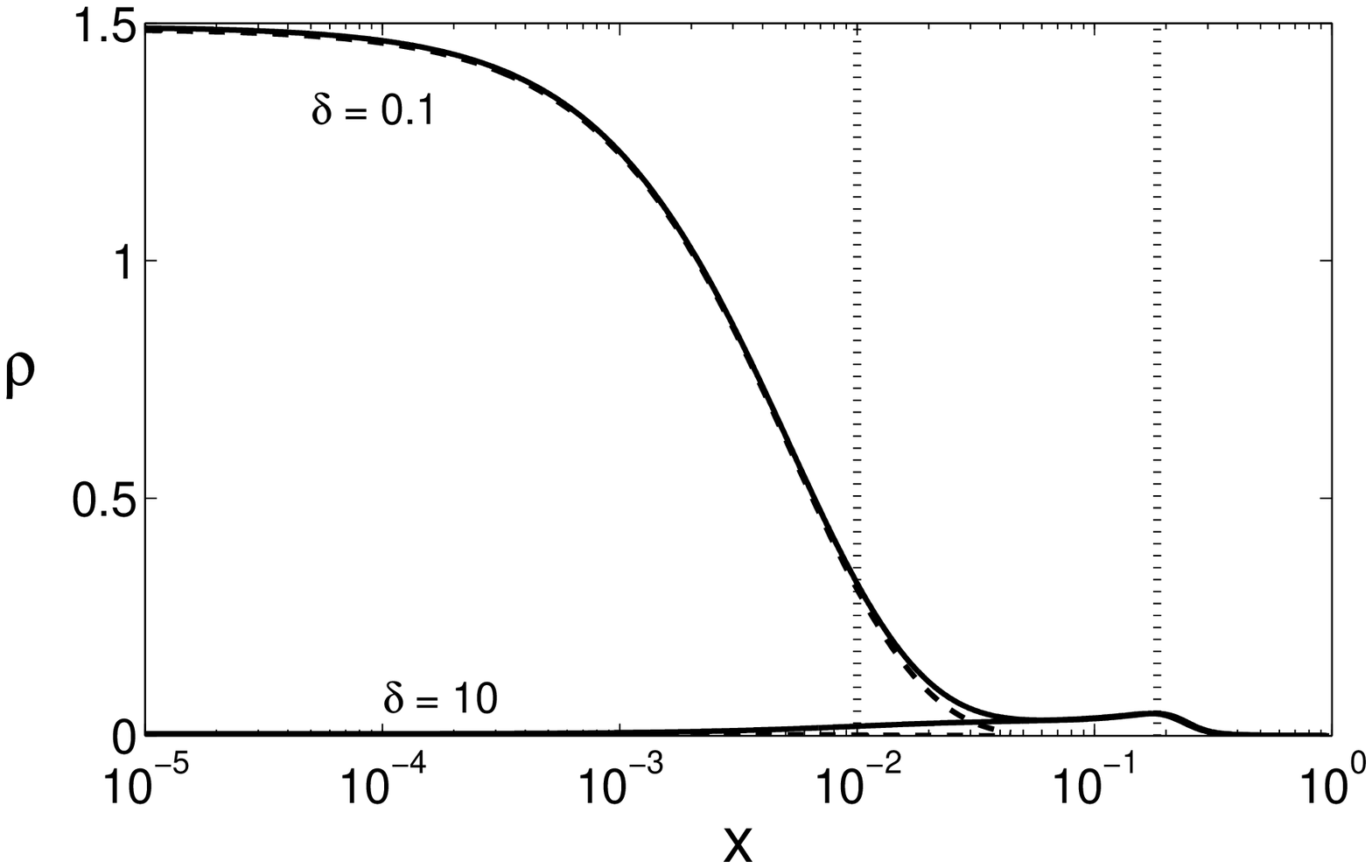}} 
\begin{minipage}[h]{5in}
\caption{
\label{figure:fields_j1_5_e0_01}
Numerical solutions (solid lines) for the dimensionless electric field
$E(x)$, average concentration $c(x)$, and charge density $\rho(x)$
above the diffusion-limited current ($\i = 1.5$) compared with leading
order asymptotic approximations (dashed lines) for $k_c = 1$, $\ir =
2$, $\epsilon = 0.01$, and $\delta = 0.1, 10$.  The leading order bulk
approximations are given by Eqs.~(\ref{eq:bulk_soln_above_lim_cur}).
In the space-charge layer, the leading order electric field is given
by Eq.~(\ref{eq:E_space_charge}), and leading order concentration is
$0$.  Finally, Eqs.~(\ref{eq:diffuse_soln_above_lim_cur_E}) and
(\ref{eq:diffuse_soln_above_lim_cur_c_plus}) are the diffuse layer
asymptotic approximations for the electric field and concentration,
respectively.  For reference, the vertical lines show where $x =
\epsilon$ and $x = x_o$.  }
\end{minipage}
\ec
\end{figure}

In the bulk, the solution remains unchanged except that $\b{c}_o$ 
cannot be approximated by $1-\i$; the contribution from the integral term 
is no longer negligible.  
The need for this correction arises from the high electric fields
required to drive current through the electrically charged space-charge 
layer.  With this minor modification, we find that the bulk solution is 
\bea
  \cb(x) &=& \b{c}_o + 2\i x \nonumber \\
  \Eb(x) &=& \frac{1}{x_o - x} 
  \label{eq:bulk_soln_above_lim_cur}
\eea
where $x_o \equiv -\b{c}_o/2\i$ is the point where the bulk concentration
vanishes (see Figure~\ref{figure:fields_j1_5_e0_01}).

\begin{figure}
\bc
\scalebox{0.45}{\includegraphics{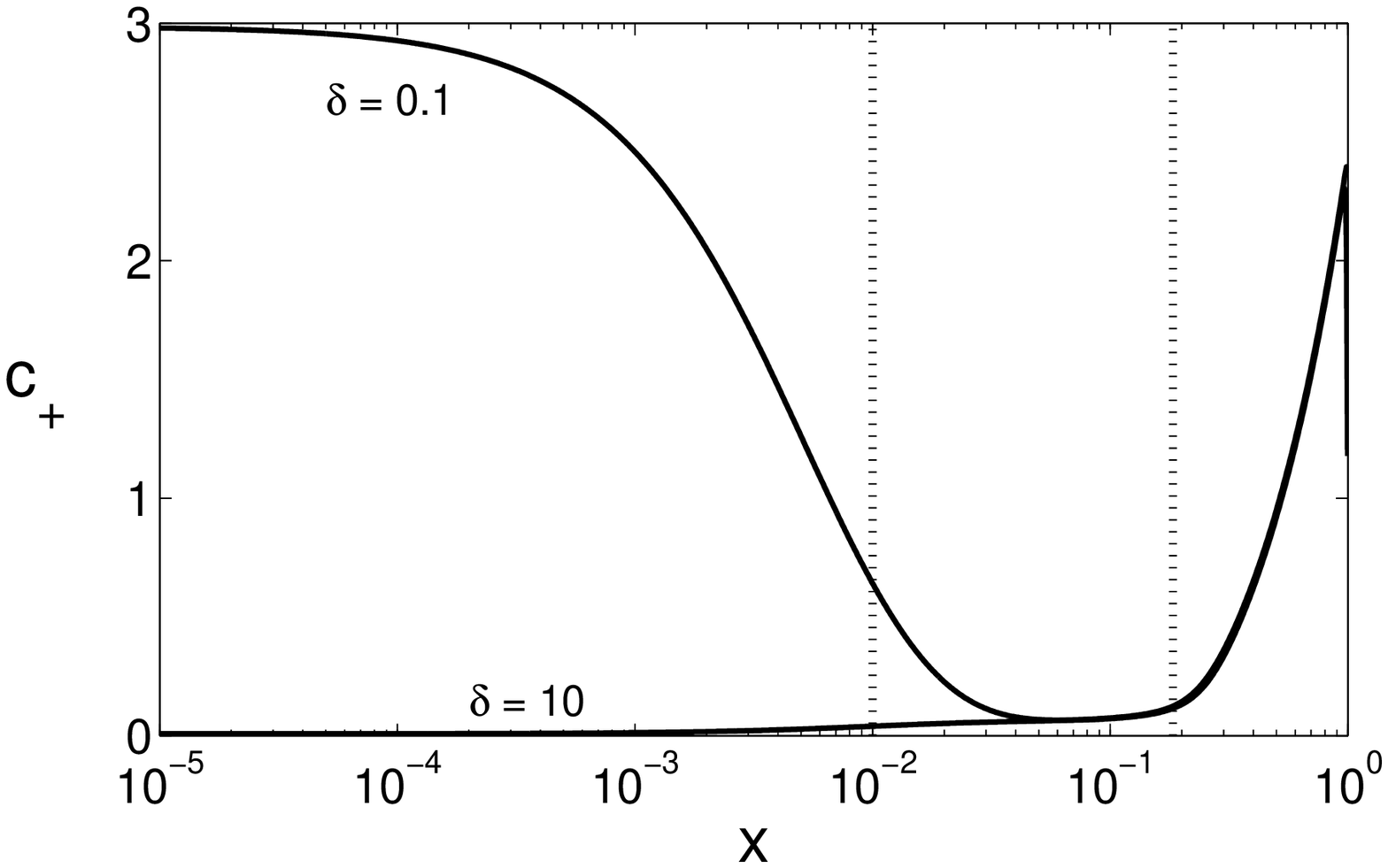}} \\
\scalebox{0.45}{\includegraphics{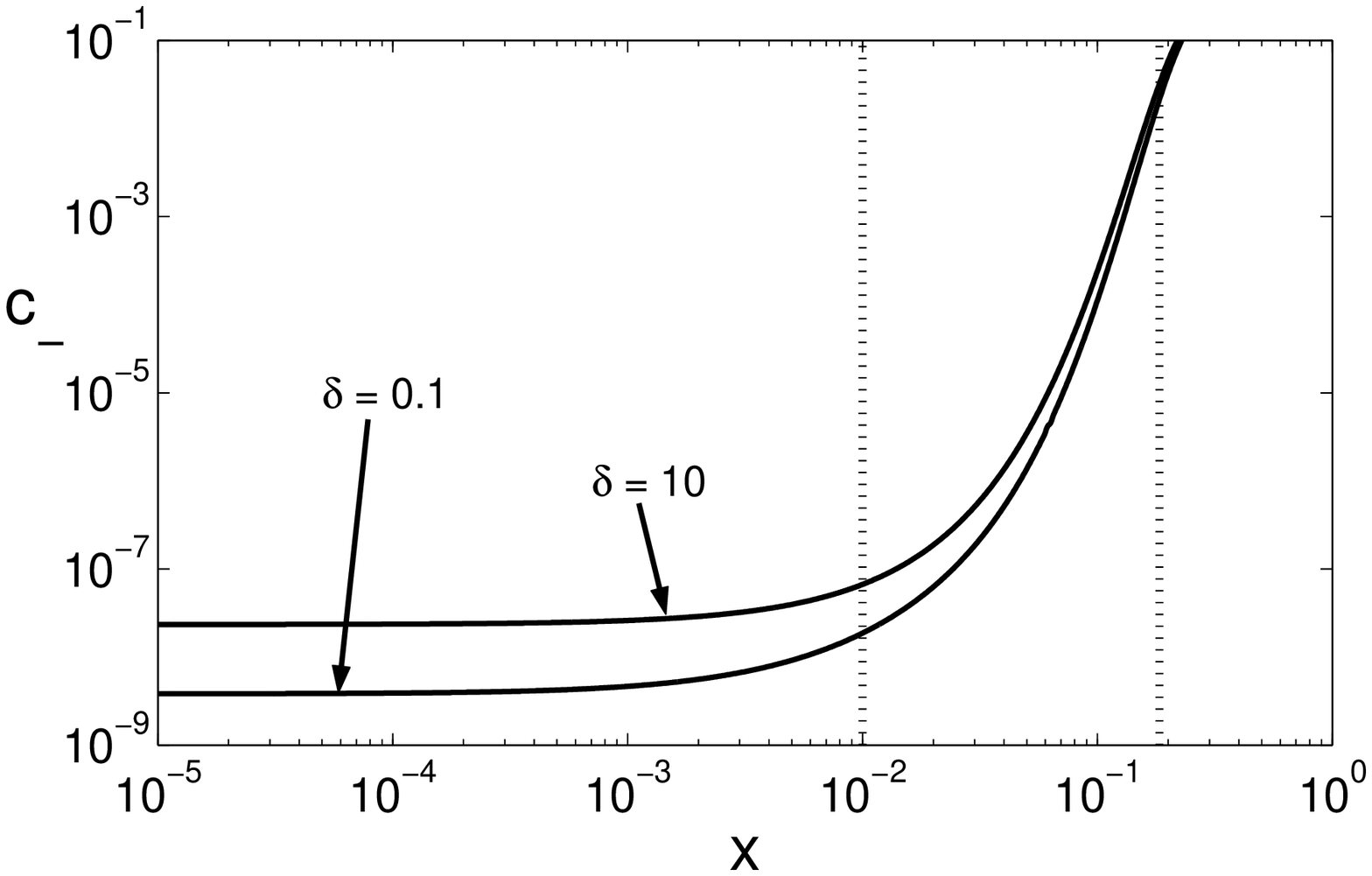}} 
\begin{minipage}[h]{5in}
\caption{
\label{figure:cation_anion_j1_5_e0_01}
Numerical solutions for the dimensionless cation and anion
concentrations above the diffusion-limited current ($\i = 1.5$) for
$k_c = 1$, $\ir = 2$, $\epsilon = 0.01$, and $\delta = 0.1, 10$.  For
reference, the vertical lines show where $x = \epsilon$ and $x = x_o$.
}
\end{minipage}
\ec
\end{figure}

\begin{figure}
\bc
\scalebox{0.34}{\includegraphics{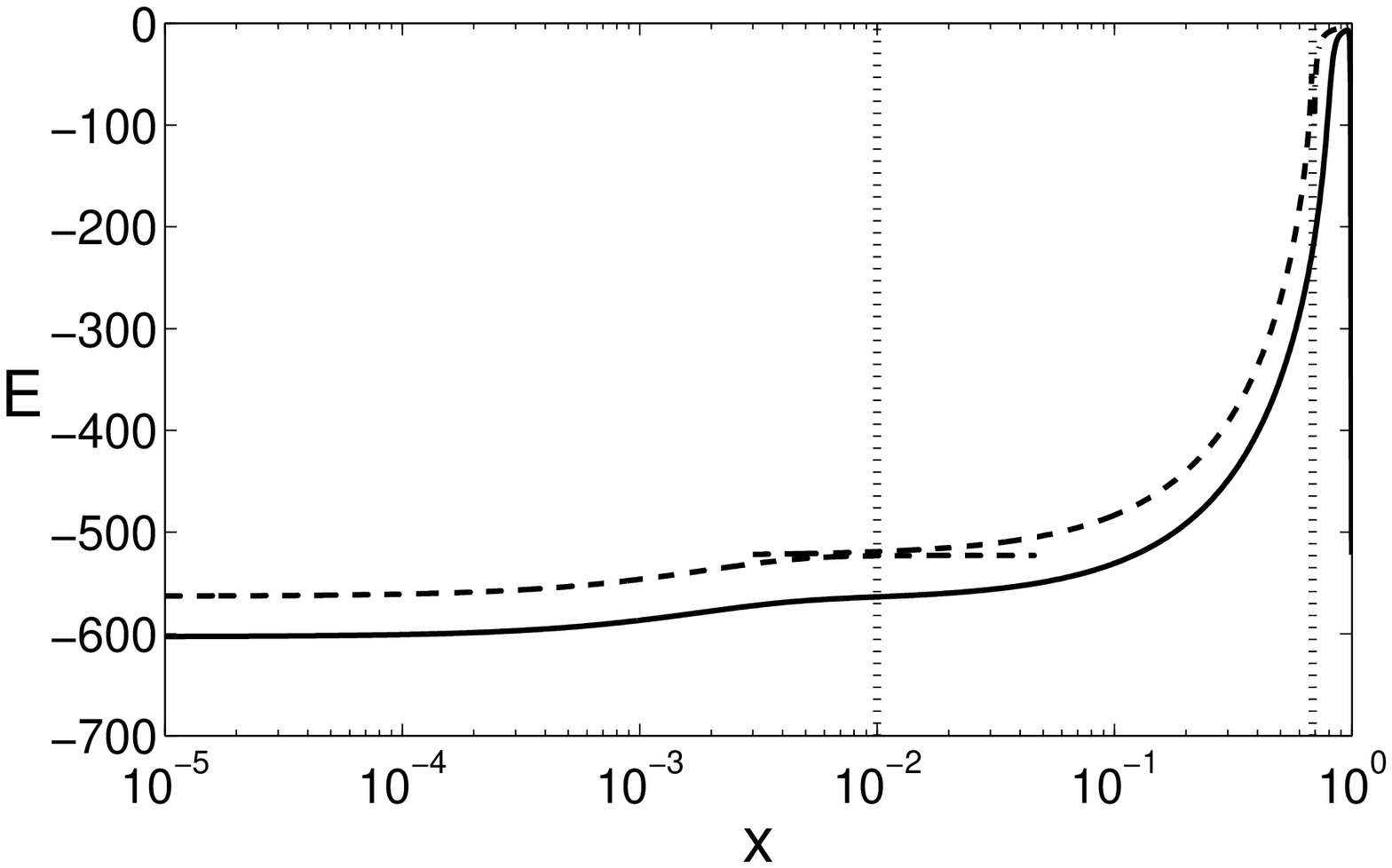}} 
\scalebox{0.34}{\includegraphics{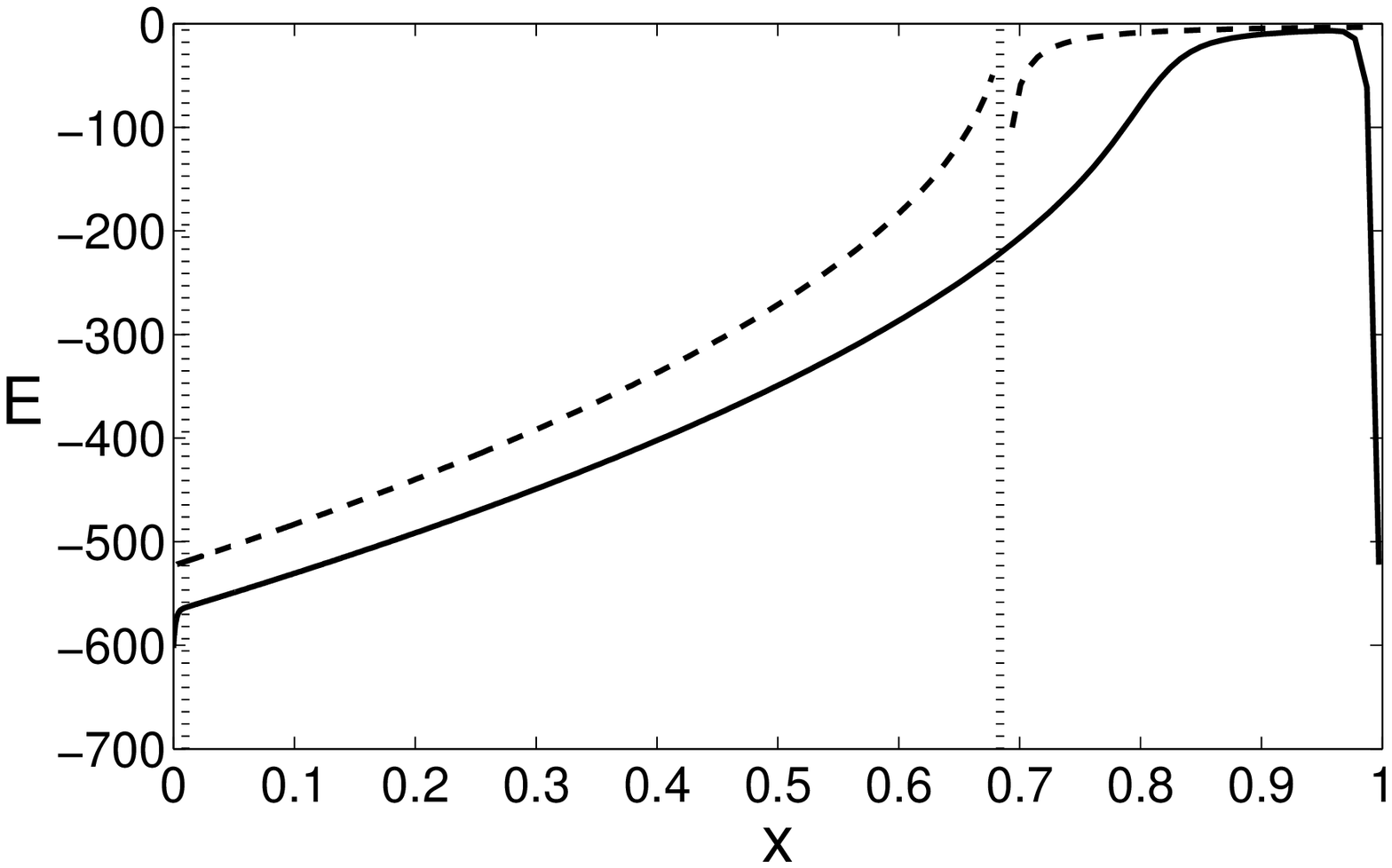}} \\
\scalebox{0.34}{\includegraphics{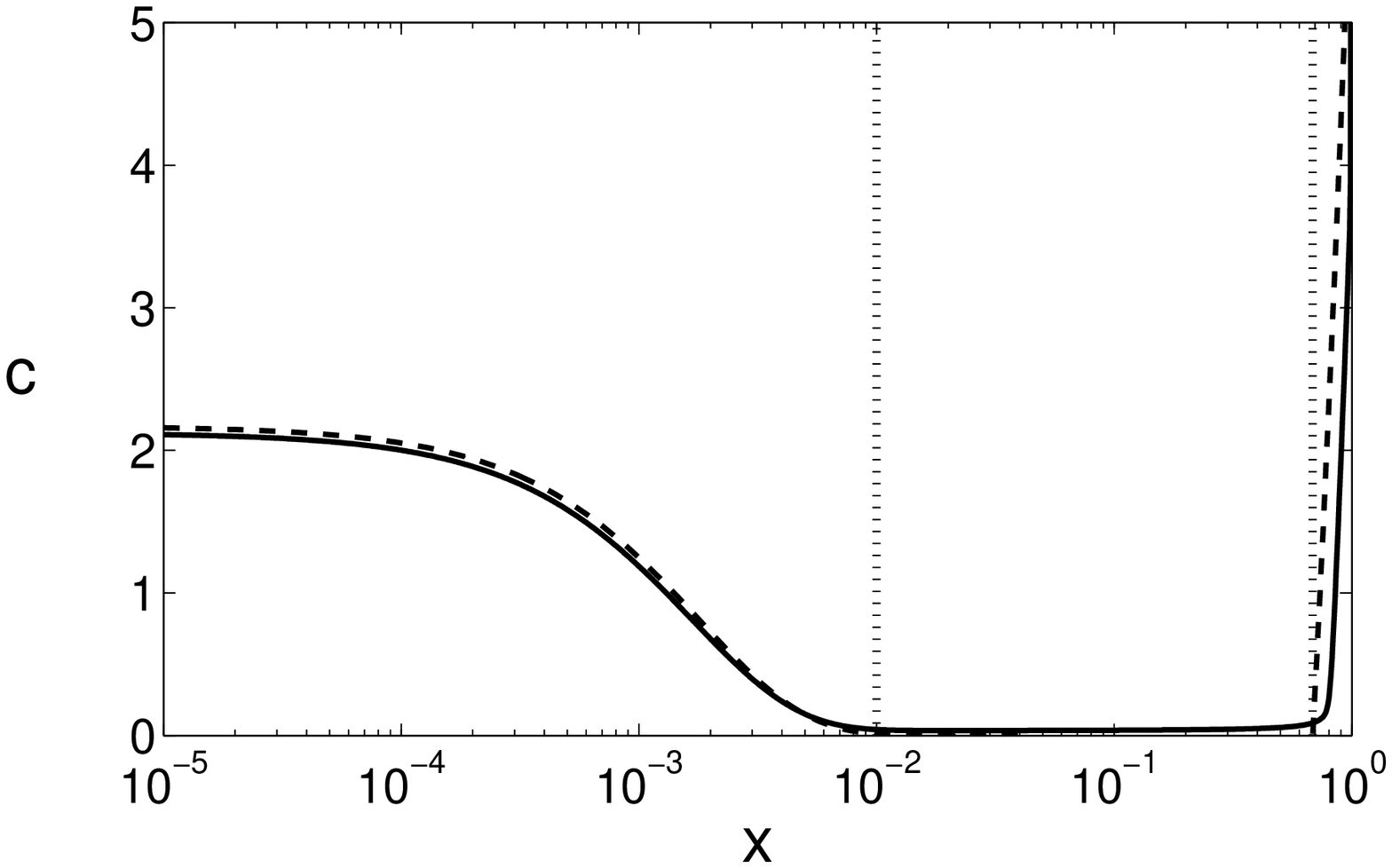}} 
\scalebox{0.34}{\includegraphics{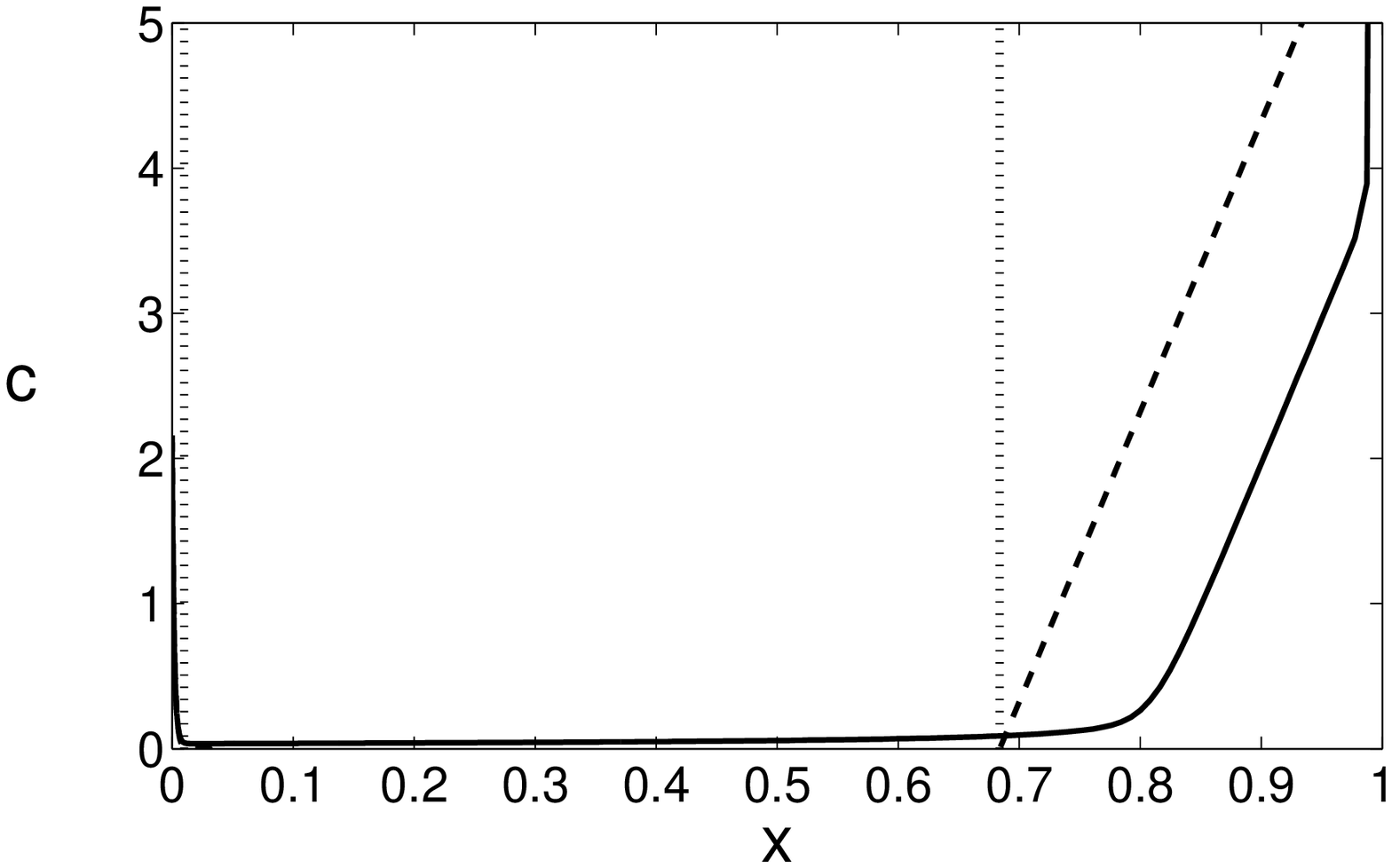}} \\
\scalebox{0.34}{\includegraphics{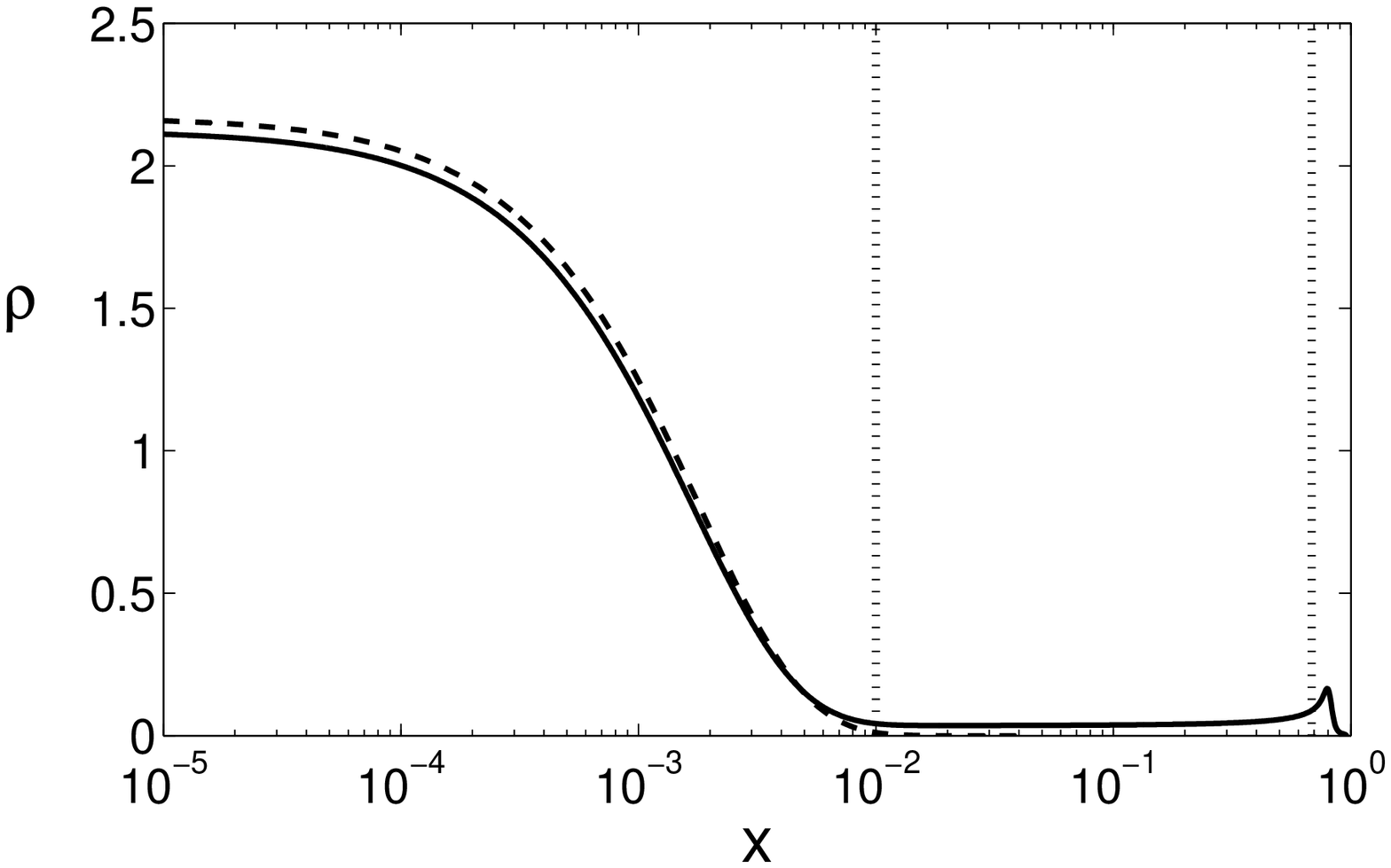}} 
\scalebox{0.34}{\includegraphics{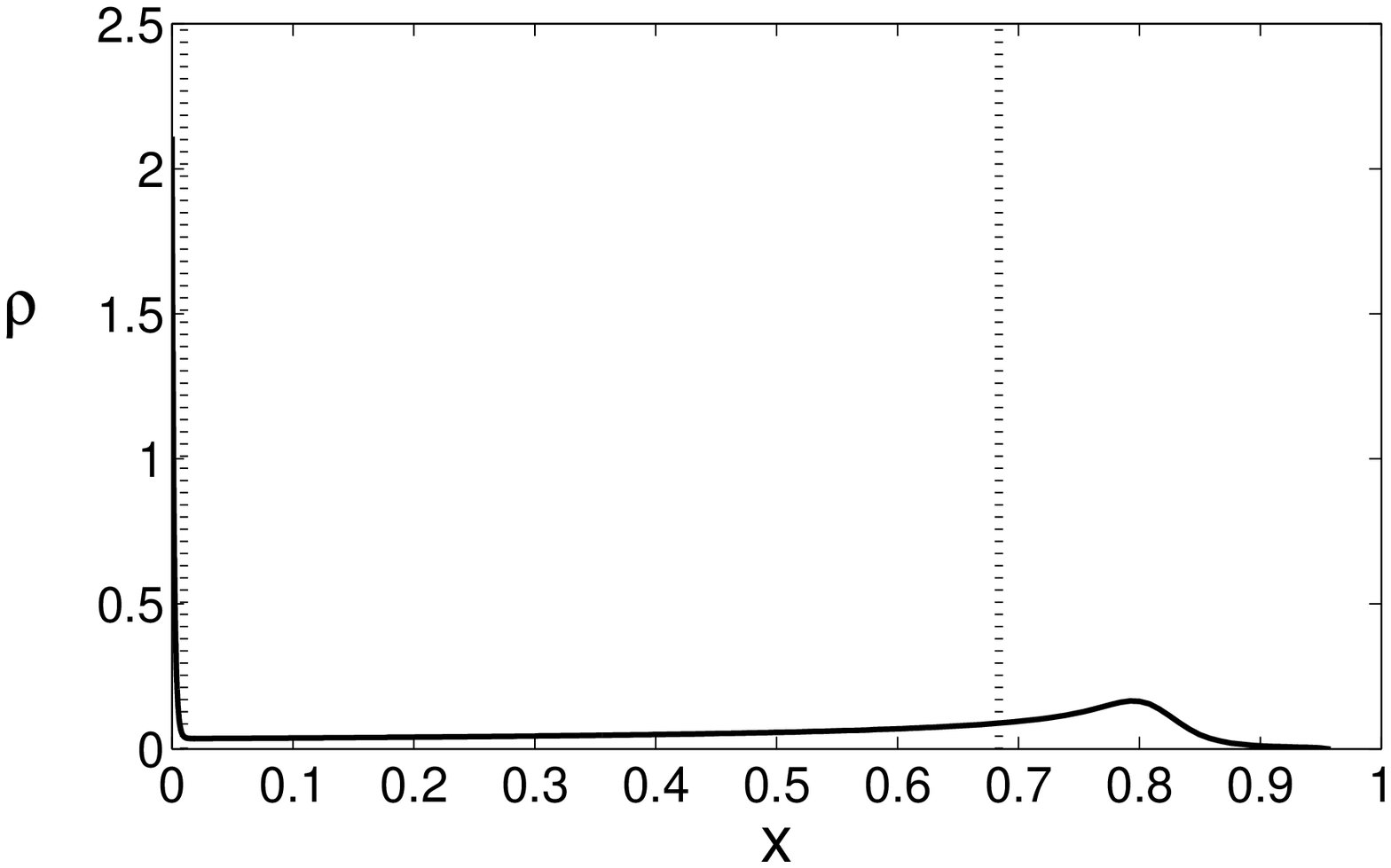}} 
\begin{minipage}[h]{5in}
\caption{
\label{figure:fields_j10_0_e0_01}
Numerical solutions (solid lines) for the dimensionless electric field
$E(x)$, average concentration $c(x)$, and charge density $\rho(x)$ far
above the diffusion-limited current ($\i = 10.0$) compared with
leading order asymptotic approximations (dashed lines) for $k_c = 1$,
$\ir = 2$, $\epsilon = 0.01$, and $\delta = 0.1$.  Each field is shown
twice: (1) with $x$ on $\log$ scale to focus on the cathode region and
(2) with $x$ on a linear scale to emphasize the interior of the cell.
Note that $\i \epsilon = 0.1$, so the asymptotic approximations are
not as good as at lower current densities.  For reference, the
vertical lines show where $x = \epsilon$ and $x = x_o$.  }
\end{minipage}
\ec
\end{figure}

In between the two $O(1)$ layers, there is a small transition layer.
Rescaling the master equation using the change of variables 
$z = (x-x_o)/\e23$ and $\Ea(z) = \e23\Eb(x)$, we 
again obtain the Smyrl-Newman equation, Eq.~(\ref{eq:SN_eqn}), 
with right hand side equal to zero.  As before, we find that the 
solution in the transition layer approaches $-1/z$ as 
$z \rightarrow \infty$.  In the other direction as $z \rightarrow -\infty$, 
we will find that the appropriate boundary condition 
is $\Ea \rightarrow -2 \sqrt{\i |z|}$ to match the electric field 
in space-charge layer.

\subsection{ Structure of the Space-Charge Layer}
Physically, we could argue that the concentration of ions in the space-charge 
layer is very small ({\it i.e.} zero at leading order) because the layer is 
essentially the result of stretching the ionic content of the overlap 
between the inner diffuse and Smyrl-Newman layers, which is small to begin 
with, over an $O(1)$ region.  This physical intuition is confirmed by
the numerical solutions shown in 
Figures~\ref{figure:fields_j1_5_e0_01}, \ref{figure:cation_anion_j1_5_e0_01},
and \ref{figure:fields_j10_0_e0_01}.
Therefore, using Eq.~(\ref{eq:c_master_eqn}), we obtain the leading 
order solution for the electric field 
\beq
  \Et \sim \frac{-2 \sqrt{\i \left ( x_o - x \right )}}{\epsilon}.
  \label{eq:E_space_charge}
\eeq
Note that the magnitude of the field is exactly what is required to 
make the integral term in $\b{c}_o$ an $O(1)$ contribution.
From this formula, it is easy to compute the charge density in the 
space-charge layer
\beq
  \rhot = \epsilon^2\frac{d \Et}{dx} \sim 
    \epsilon \sqrt{\frac{\i}{x_o - x}},
  \label{eq:rho_space_charge}
\eeq
which is an order of magnitude larger than the $O(\epsilon^2)$ charge
density in the bulk.  The $O(\epsilon)$ charge density also implies
that the concentration must be at least $O(\epsilon)$ because the
anion concentration, $c-\rho$, is positive.

With the electric field given by Eq.~(\ref{eq:E_space_charge}), 
we can determine the values of $x_o$ and $\b{c}_o$ by solving the 
system of equations given by the definition of $x_o$ and $\b{c}_o$.
Using Eq.~(\ref{eq:c_o_master}) to calculate $\b{c}_o$ and noticing
that the leading order contribution to the integral comes from the 
space-charge layer, we obtain
\beq
  \b{c}_o \sim 1 - \i ( 1 + x_o^2 ).
\eeq
Combining this result with $x_o = -\b{c}_o/2\i$, we find that
\bea
  x_o \sim 1 - \i^{-1/2} \ \ \ , \ \ \ 
  \b{c}_o \sim 2 \left (\i^{1/2} -\i \right),
\eea
which can be substituted into Eqs.~(\ref{eq:bulk_soln_above_lim_cur})
and (\ref{eq:E_space_charge}) to yield the leading order solutions in
the bulk and space-charge layers.  It should be noted that the
expression for $x_o$ is consistent with the estimate for the width
found by Bruinsma and Alexander~\cite{bruinsma1990} and
Chazaviel~\cite{chazalviel1990} in the limits $\i-1 \ll 1$ and small
space-charge layer ($x_o \ll 1$), although our analysis also applies to
much larger voltages.

The results obtained via physical arguments in the previous few paragraphs
motivate an asymptotic series expansion for $E$ whose leading order term 
is $O(1/\epsilon)$.  Moreover, because we want to be able to balance the
current density at second-order, we choose the second-order term to be 
$O(\i)$.  Thus, we have 
\beq
  \Et = \frac{1}{\epsilon} E_{-1} + E_0 \i + \ldots.
  \label{eq:E_space_charge_asym_series}
\eeq 
Note that in this asymptotic series, the first term only dominates the
second term as long as $\i \ll 1/\epsilon$, so the following analysis only
holds for current densities far below $O(1/\epsilon)$.  
Figure~\ref{figure:fields_j10_0_e0_01} illustrates the breakdown
of the leading-order asymptotic solutions at very high current densities.
While the qualitative features of the asymptotic approximation are correct
({\it e.g.} shape of $E(x)$ in the diffuse layer and slope of $c(x)$ 
in the bulk), the quality of the approximation is clearly less than 
at lower values of $\i$.

The key advantage of a more systematic asymptotic analysis is that we are able 
to calculate the leading-order behavior of the space-charge layer 
concentration $\ct$, which is not possible with only knowledge of the 
leading order behavior for the electric field.  
Substituting Eq.~(\ref{eq:E_space_charge_asym_series})
into the master equation (\ref{eq:E_master_eqn}), it is straightforward to
obtain
\bea
  \Et \sim - \frac{2} {\epsilon}\sqrt{\i \left ( x_o - x \right )}
  - \frac{1}{2(x_o - x)} + \ldots.
  \label{eq:E_space_charge_first_two_terms}
\eea
Using this expression in Eq.~(\ref{eq:c_master_eqn}), we find the dominant 
contribution to $\ct$ is exactly the same as $\rhot$:
\beq
  \ct \sim \epsilon \sqrt{\frac{\i}{x_o - x}}.
  \label{eq:c_space_charge}
\eeq
Since $c_- = c - \rho$, this result leads to an important physical
conclusion --- {\it The space-charge layer is essentially depleted of
anions}, $c_- = o(\epsilon)$, as is clearly seen in
Figures~\ref{figure:fields_j1_5_e0_01} and
\ref{figure:cation_anion_j1_5_e0_01}. This contradicts our
macroscopic intuition about electrolytes, but, in very thin films,
complete anion deplection might occur. For example, in a microbattery
developed for on-chip power sources using the Li/SiO$_2$/Si system,
lithium ion conduction has recently been demonstrated in nano-scale
films of silicon oxide, where there should not be any counter ions or
excess electrons~\cite{nava}.

At leading order as $\epsilon\rightarrow 0$, the anion concentration,
$c_-$, can be set to zero in the space-charge layer, leaving the
following two governing equations:
\bea
\frac{dc_+}{dx} + c_+ \frac{d\phi}{dx} & = & 4 \i
  \label{eq:c+eq_space_charge}\\
- \epsilon^2 \frac{d^2\phi}{dx^2} &=& \frac{1}{2}c_+. 
  \label{eq:phieq_space_charge}
\eea
As with binary electrolyte case, these equations can be reduced to a single
equation for the electric potential: 
\bea
  \frac{d^3 \phi}{dx^3} + \frac{d^2 \phi}{dx^2} \frac{d \phi}{dx} = 
  -\frac{2\i}{\epsilon^2}.
  \label{eq:phi_master_eq_space_charge}
\eea
Integrating this equation once, we obtain a Riccati equation for 
$\frac{d\phi}{dx}$
\bea
  \frac{d^2 \phi}{dx^2} + \frac{1}{2} \left ( \frac{d\phi}{dx} \right )^2 
  = -\frac{2\i}{\epsilon^2}(x-x_o) + h,
  \label{eq:phi_riccati}
\eea
where $h$ is an integration constant.  Using the transformations
\bea
  u \equiv e^{\phi/2} \ \ \ , \ \ \ 
  z \equiv - \frac{\i^{1/3}}{\epsilon^{2/3}} \left ( x - x_o \right ) 
    + \frac{\epsilon^{4/3} h}{2 \i^{2/3}},
\eea
we find that $u$ satifies Airy's equation 
\beq
  \frac{d^2 u}{dz^2} - z u = 0.
\eeq
Thus, the general solution for $\phi(x)$ is 
\beq
  \phi(x) = 2 \log \left [
    a_1 Ai\left (\frac{\i^{1/3}}{\epsilon^{2/3}} \left ( x_o - x\right ) 
    + \beta h \right )
  + a_2 Bi\left (\frac{\i^{1/3}}{\epsilon^{2/3}} \left (x_o - x\right ) 
    + \beta h \right )
  \right ],  \label{eq:Airysoln}
\eeq
where $a_1$ and $a_2$ are constants determined by boundary conditions
and $\beta = \frac{\epsilon^{4/3}}{2 \i^{2/3}}$.

To simplify this expression, note that in the limit 
$\epsilon \rightarrow 0$, the potential drop between 
$x=x_o$ and $x=0$ is approximately
\beq
  \phi(x_o) - \phi(0) \sim 
  2 \log \left [
    \frac{a_1 Ai(0) + a_2 Bi(0)}
    {a_1 Ai \left (\frac{x_o \i^{1/3}}{\epsilon^{2/3}} \right ) 
     + a_2 Bi \left (\frac{x_o \i^{1/3}}{\epsilon^{2/3}} \right )}
  \right ].
\eeq
Now, using the large argument behavior of the Airy functions, we see that
as $\epsilon \rightarrow 0$, the argument of the logarithm approaches
zero.  Thus, we are lead to the conclusion that the electric potential 
at $x = x_o$ is less than at $x = 0$.  But, this is completely inconsistent
with our physical intuition and the numerical results, which show that 
$\phi(x_o) - \phi(0) > 0$.  Therefore, it must be the case that 
$a_2 \approx 0$ so that
\beq
  \phi(x) = 2 \log \left [
    a_1 Ai\left (\frac{\i^{1/3}}{\epsilon^{2/3}} \left ( x_o - x\right ) 
    + \beta h \right )
  \right ]
\eeq
and
\beq
  E(x) = \frac{2 \i^{1/3}}{\epsilon^{2/3}} 
    \frac{Ai'\left (\frac{\i^{1/3}}{\epsilon^{2/3}} \left ( x_o - x\right ) 
    + \beta h \right )}
    {Ai\left (\frac{\i^{1/3}}{\epsilon^{2/3}} \left ( x_o - x\right ) 
    + \beta h \right )}.
\eeq
Finally, by using the asymptotic form of $Ai(z)$ and $Ai'(z)$ as 
$z \rightarrow \infty$,
we find that in the $\epsilon \rightarrow 0$ limit, the leading order
approximation for the electric field when the region is depleted of
anions is exactly Eq.~(\ref{eq:E_space_charge}).  

That equivalence of the single-ion equations and the full governing
equations at leading-order mathematically confirms the physically 
interpretation of the space-charge layer as a region of anion depletion.
From an alternative perspective, it also reminds us that under extreme
conditions, it may be necessary to rethink our assumptions about what
physical effects are dominant. 

\subsection{ Boundary Layers Above the Limiting Current}
To complete our analysis of the high-current regime, $1 + O(\e23) \ll
\i \ll O(1/\epsilon)$, we must consider the boundary layers.  At
the anode, all fields are $O(1)$, so we recover the usual Gouy-Chapman
solution with the minor modification that $c_1 = 2\sqrt{\i}$ which is
the value $\cb$ takes as $x \rightarrow 1$.  The cathode structure,
however, is much more interesting because it is depleted of anions
(see Figure~\ref{figure:cation_anion_j1_5_e0_01}). To our knowledge,
this non-equilibrium inner boundary layer on the space-charge region,
related to the reaction boundary condition at the cathode, has not
been analyzed before.

As in the space-charge layer, the leading-order governing equations in 
this layer are those of a single ionic species with no counterions 
Eqns.~(\ref{eq:c+eq_space_charge}) and (\ref{eq:phieq_space_charge}).
Rescaling those equations using $x = \epsilon y$, we obtain 
\bea
\frac{d\cc_+}{dy} + \cc_+ \frac{d\phic}{dy} &=& 4\i\epsilon \approx 0  
    \label{eq:single_ion_eqns_NP} \\
- \frac{d^2\phic}{dy^2} &=& \frac{1}{2}\cc_+ 
    \label{eq:single_ion_eqns_poisson}.
\eea
From these equations, it is immediately clear that the cations have a
Boltzmann equilibrium profile at leading order: $c_+ \propto
e^{-\phi(y)}$.  As in the analysis for the space-charge layer, it is
possible to find a general solution to
Eqns.~(\ref{eq:single_ion_eqns_NP}) and
(\ref{eq:single_ion_eqns_poisson}).  By combining these equations
and integrating, we find that the potential in the cathode boundary layer
has the form
\beq
  \phic \sim \log 
  \left [ 
    \sinh^2 \left (p y + q \right ) 
  \right ] + r,
\eeq
where $p$, $q$, and $r$ are integration constants. 
Therefore, the electric field and concentration are
\bea 
  \Ec(y) &\sim& - 2 p
  \coth \left (p y + q \right ) 
  \label{eq:diffuse_soln_above_lim_cur_E} \\
  \cc(y) &=& \frac{1}{2}\cc_+(y) \sim \frac{2 p^2}
  {\sinh^2 \left (p y + q \right )}
  \label{eq:diffuse_soln_above_lim_cur_c_plus}
\eea

Matching the electric fields in the diffuse and space-charge layers, 
we find that $p \sim \sqrt{\i x_o}$.  
Note that because $p = O(\sqrt{\i})$, the electric field in the diffuse 
charge layer is $O(\sqrt{\i}/\epsilon)$ which is same order of magnitude 
as in the space-charge layer.
To solve for $q$, we use the expression for $p$ in the cathode Stern and 
Butler-Volmer boundary conditions, which leads to the following nonlinear 
equation: 
\beq
  \frac{4 k_c \i x_o}{\sinh^2 q} 
  \exp \left ( 2 \alpha_c \delta \sqrt{\i x_o} \coth q \right )
  - \ir \exp \left ( - 2 \alpha_a \delta \sqrt{\i x_o} \coth q \right )  
  = \i.
\eeq
In the limit of small $\delta$, we can use fixed-point iteration to 
obtain an approximate solution
\beq
  q \sim \sinh^{-1} \left( 
  2 \sqrt {\frac{k_c \i x_o 
     \exp\left (2 \alpha_c \delta \sqrt{\i x_o} \coth q_o \right )}
       {\i + \ir 
     \exp \left (-2 \alpha_a \delta \sqrt{\i x_o} \coth q_o \right) } } 
  \right)
\eeq
where $q_o$ has the same form as $q$ with $(\coth q_o)$ set equal to 1.
For $\delta \gg 1$, the leading order equation is
\beq
  \frac{4 k_c \i x_o}{\sinh^2 q} 
  \exp \left ( 2 \alpha_c \delta \sqrt{\i x_o} \coth q \right )
  \sim \i,
\eeq
which implies that $q \gg 1$ so that the left-hand side can be small enough 
to balance the current.  
Thus, by using $\coth q \approx 1$ and $\sinh q \approx \exp(q)/2$, we find 
that $q \sim \alpha_c \delta \sqrt{\i x_o} + \frac{1}{2} \log(16 k_c x_o)$.
The agreement of these asymptotic approximations with the exact solutions
in the diffuse charge layer is illustrated in 
Figure~\ref{figure:fields_j1_5_e0_01}.

\section{Polarographic Curves} 
We are now in a position to compute the leading-order behavior of 
the polarographic curve at and above the classical limiting current. 
Recall that the formula for the cell voltage is given by 
\beq
  v =  -\delta \epsilon E(0) + \int_0^1 -E(x) dx - \delta \epsilon E(1).
\eeq
The integral is the voltage drop through the interior of the cell
and the first and last terms account for the potential drop across the 
Stern layers.  

\begin{table}
\begin{center}
\caption{
\label{table:cell_voltages}
\noindent Comparison of the asymptotic approximations 
Eqs.~(\ref{eq:v_vs_i_lim_cur}) and 
(\ref{eq:v_vs_i_above_lim_cur}) with numerically calculated
values for the cell voltage at various $\epsilon$ and $\delta$ values.
These cell voltages were computed with $k_c = 1$ and $\ir = 2$.
}
\begin{tabular}{ccc}
\hline \hline & & \\
\begin{tabular}{c|r|rr}
\multicolumn{4}{c}{$\i$ = 1.0} \\
\hline
$\epsilon$ & $\delta$ & $v_{exact}$ & $v_{asym}$ \\
\hline \hline
1e-4 & 0.01 & 13.125 & 12.101 \\ \hline
1e-4 & 1.00 & 13.222 & 12.374 \\ \hline
1e-4 & 10.0 & 14.290 & 13.571 \\ \hline
1e-3 & 0.01 & 10.165 & 9.146 \\ \hline
1e-3 & 1.00 & 10.277 & 9.475 \\ \hline
1e-3 & 10.0 & 11.552 & 10.890 \\ \hline
1e-2 & 0.01 & 7.339 & 6.303 \\ \hline
1e-2 & 1.00 & 7.479 & 6.729 \\ \hline
1e-2 & 10.0 & 9.228 & 8.465 \\ \hline
1e-1 & 0.01 & 4.922 & 3.649 \\ \hline
1e-1 & 1.00 & 5.005 & 4.219 \\ \hline
1e-1 & 10.0 & 7.995 & 6.327 \\ \hline

\end{tabular}
& 	&
\begin{tabular}{c|r|rr}
\multicolumn{4}{c}{$\i$ = 1.5} \\
\hline
$\epsilon$ & $\delta$ & $v_{exact}$ & $v_{asym}$ \\
\hline \hline
1e-4 & 0.01 & 1297.799 & 1289.621 \\ \hline
1e-4 & 1.00 & 1297.048 & 1291.101 \\ \hline
1e-4 & 10.0 & 1305.318 & 1300.129 \\ \hline
1e-3 & 0.01 & 140.207 & 132.790 \\ \hline
1e-3 & 1.00 & 139.450 & 134.270 \\ \hline
1e-3 & 10.0 & 147.717 & 143.299 \\ \hline
1e-2 & 0.01 & 22.434 & 15.725 \\ \hline
1e-2 & 1.00 & 21.624 & 17.206 \\ \hline
1e-2 & 10.0 & 29.886 & 26.234 \\ \hline
1e-1 & 0.01 & 9.479 & 2.637 \\ \hline
1e-1 & 1.00 & 7.790 & 4.118 \\ \hline
1e-1 & 10.0 & 16.088 & 13.146 \\ \hline

\end{tabular} \\
& & \\
\hline \hline
\end{tabular}
\end{center}
\end{table}
At the limiting current, $\i=1$, we can estimate the voltage drop across the 
cell by using the bulk and diffuse layer electric field to approximate the 
field in the Smyrl-Newman transition layer to obtain
\bea
   v &\sim& - \delta \epsilon E(0) 
        + \int_0^{\e23} -E(x) dx + \int_{\e23}^1 -E(x) dx 
        - \delta \epsilon E(1) 
       \\
    &\sim& 
      2 \frac{\delta}{b} + 
      2 \log \left ( \frac{\epsilon^{-1/3} + b}{b} \right ) 
      -\frac{2}{3}\log \epsilon.
  \label{eq:v_vs_i_lim_cur}
\eea
Notice that in the small $\delta$ limit, this expression reduces to 
$v \sim -\frac{4}{3} \ln \epsilon$ as $\epsilon \rightarrow 0$ 
which can be observed in the polarographic curves in 
Figure~6 in~\cite{part1}.  
Table \ref{table:cell_voltages} compares this approximation with the 
exact cell voltage for a few $\epsilon$ and $\delta$ values.
For small $\epsilon$ values ($\epsilon \le 0.01$), the 
asymptotic approximations are fairly good (within 5\% to 10\%).

Above the limiting current, the space-charge layer makes the 
dominant contribution to the cell voltage.  
Using Eqns.~(\ref{eq:bulk_soln_above_lim_cur}) and 
(\ref{eq:E_space_charge}) in the formula for the
cell voltage, we find that
\bea
  v &\sim& \frac{4 \sqrt{\i}}{3\epsilon} \left ( 1 - \i^{-1/2} \right )^{3/2}
   + 2\delta \left ( \i -\sqrt{\i}\right ) ^{1/2} \coth q
   - \frac{1}{2} \log \i - 2/3 \log \epsilon.
  \label{eq:v_vs_i_above_lim_cur}
\eea
The first two terms in this expression estimate the voltage drop across
the space-charge and the cathode Stern layers, respectively.  The
last two terms are the subdominant contribution from the bulk where we have 
somewhat arbitrarily taken $x = x_o + \epsilon^{2/3}$ as the boundary 
between the bulk layer and the Smyrl-Newman transition layer.  
Notice that we ignore the contribution from the cathode diffuse and 
Smyrl-Newman layers.  It is safe to neglect the diffuse layer because 
it is an $O(1)$ contribution.  However, the Smyrl-Newman layer has a 
non-neglible potential drop that we have to accept as error 
since we do not have an analytic form for the solution in that region.

\begin{figure}
\bc
\scalebox{0.5}{\includegraphics{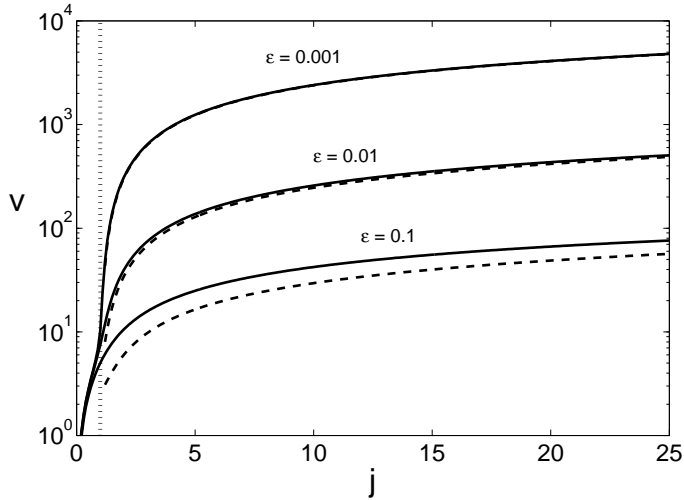}}
\begin{minipage}[h]{5in}
\caption{
\label{figure:j_vs_v_high_current}
Comparison of numerical polarographic curves (dashed lines) with
leading-order asymptotic approximations (solid lines) given in
Eq.~(\ref{eq:v_vs_i_above_lim_cur}) for several values of $\epsilon$
with $\delta = 1.0$, $k_c = 1$ and $\ir = 2$.  For $\epsilon = 0.001$,
the numerical and asymptotic polarographic curves are
indistinguishable on this graph.  For reference, the vertical, dashed
line shows the classical diffusion-limited current $\i = 1$.  }
\end{minipage}
\ec
\end{figure}

Figure~\ref{figure:j_vs_v_high_current} shows that the asymptotic 
polarographic curves are quite accurate for sufficiently small $\epsilon$ 
values.  
In Table~\ref{table:cell_voltages}, we compare the results predicted
by the asymptotic formula with numerical results for a few specific values 
of $\epsilon$ and $\delta$.  
It is interesting that the approximation is also better for large 
$\delta$ values (we explain this observation in the next section).
Also, while the $\log \epsilon$ term is subdominant, 
it makes a significant contribution to the cell voltage for 
$\epsilon$ values as small as $0.01$.  

As with the width of the space-charge layer, $x_o$, our expression for
the cell voltage, Eq.~(\ref{eq:v_vs_i_above_lim_cur}), is consistent
with the results of Bruinsma and Alexander~\cite{bruinsma1990} and
Chazaviel~\cite{chazalviel1990}, near the limiting curent, $\i
\rightarrow 1^+$, while remaining valid at much larger currents, 
$\i = O(1/\epsilon)$

\section{ Effects of the Stern-Layer Capacitance} 
The inclusion of the Stern layer in the boundary conditions allows us
to explore the effects of the intrinsic surface capacitance on the
structure of the cell.  From Figures
\ref{figure:fields_j1_0_e0_01} through \ref{figure:cation_anion_j1_5_e0_01},
we can see that smaller Stern-layer capacitances 
({\it i.e.} larger $\delta$ values)
decrease the concentration and electric field strength in the cathode
diffuse layer.  
This behavior arises primarily from the influence of the
electric field on the chemical kinetics at the electrode surfaces.
When the capacitance of the Stern layer is low, small electric fields at 
the cathode surface translate into large potential drops across the Stern
layer, Eq.~(\ref{eq:potbc0}), which help drive the deposition reaction, 
Eq.~(\ref{eq:potbvbc0}).  As a result, neither the electric field nor the 
cation concentration need to be very large at the cathode to support
high current densities.
These results confirm our physical intuition that it is only 
important to pay attention to the diffuse layer when the Stern layer 
potential drop is negligible ({\it i.e} $\delta \ll 1$).

\begin{figure}[h]
\bc
\scalebox{0.35}{\includegraphics{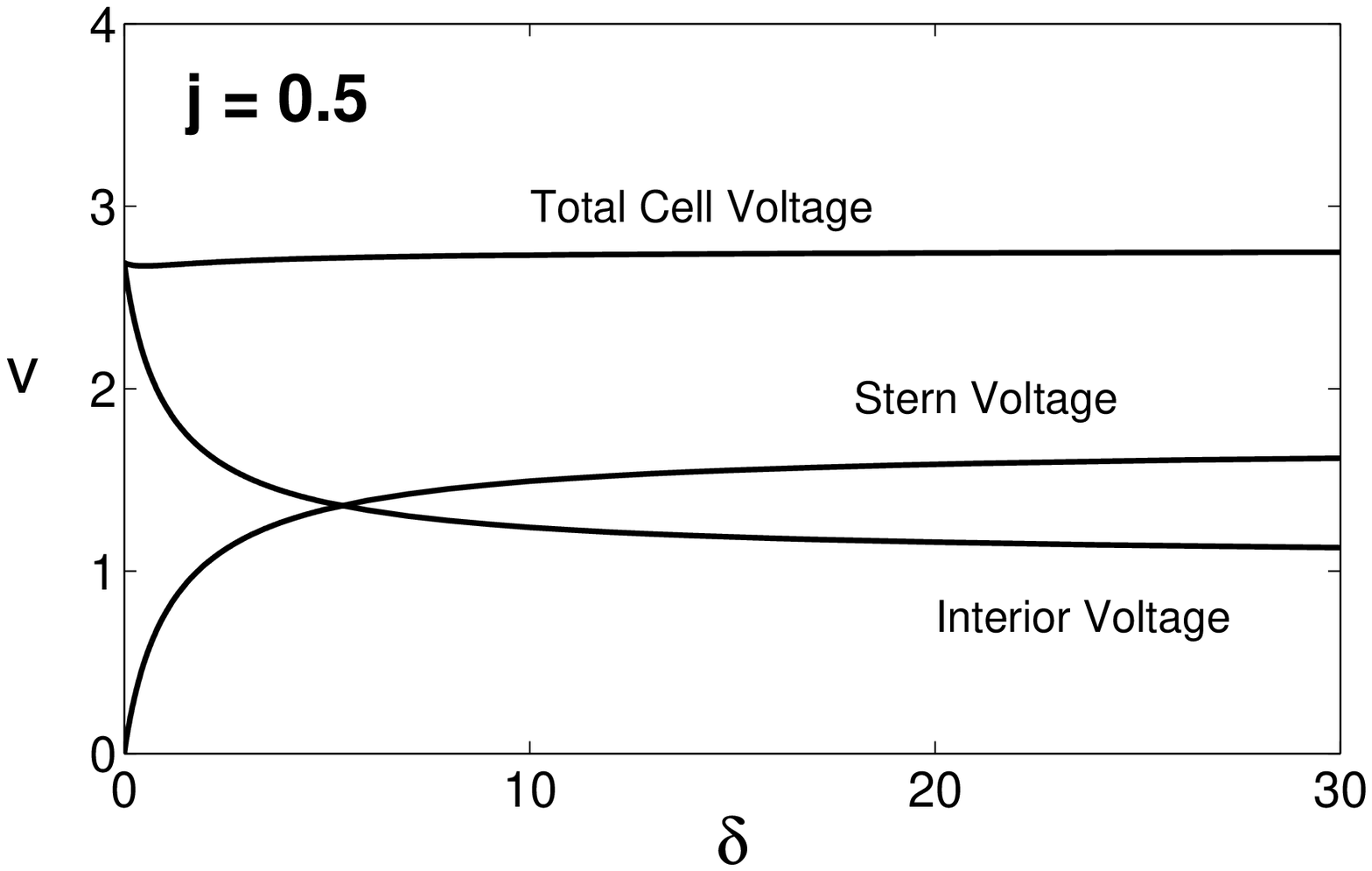}}
\scalebox{0.35}{\includegraphics{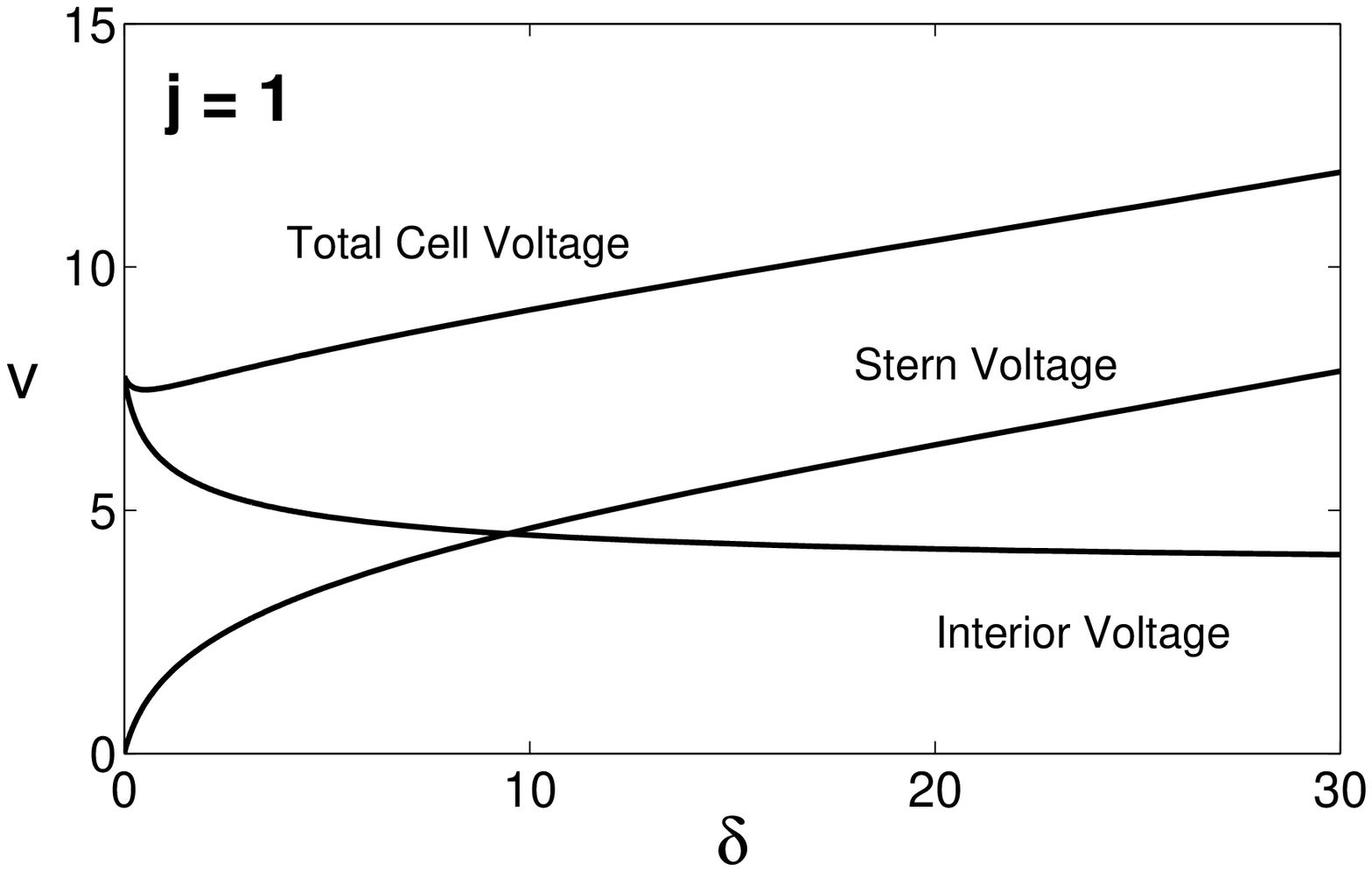}}
\scalebox{0.35}{\includegraphics{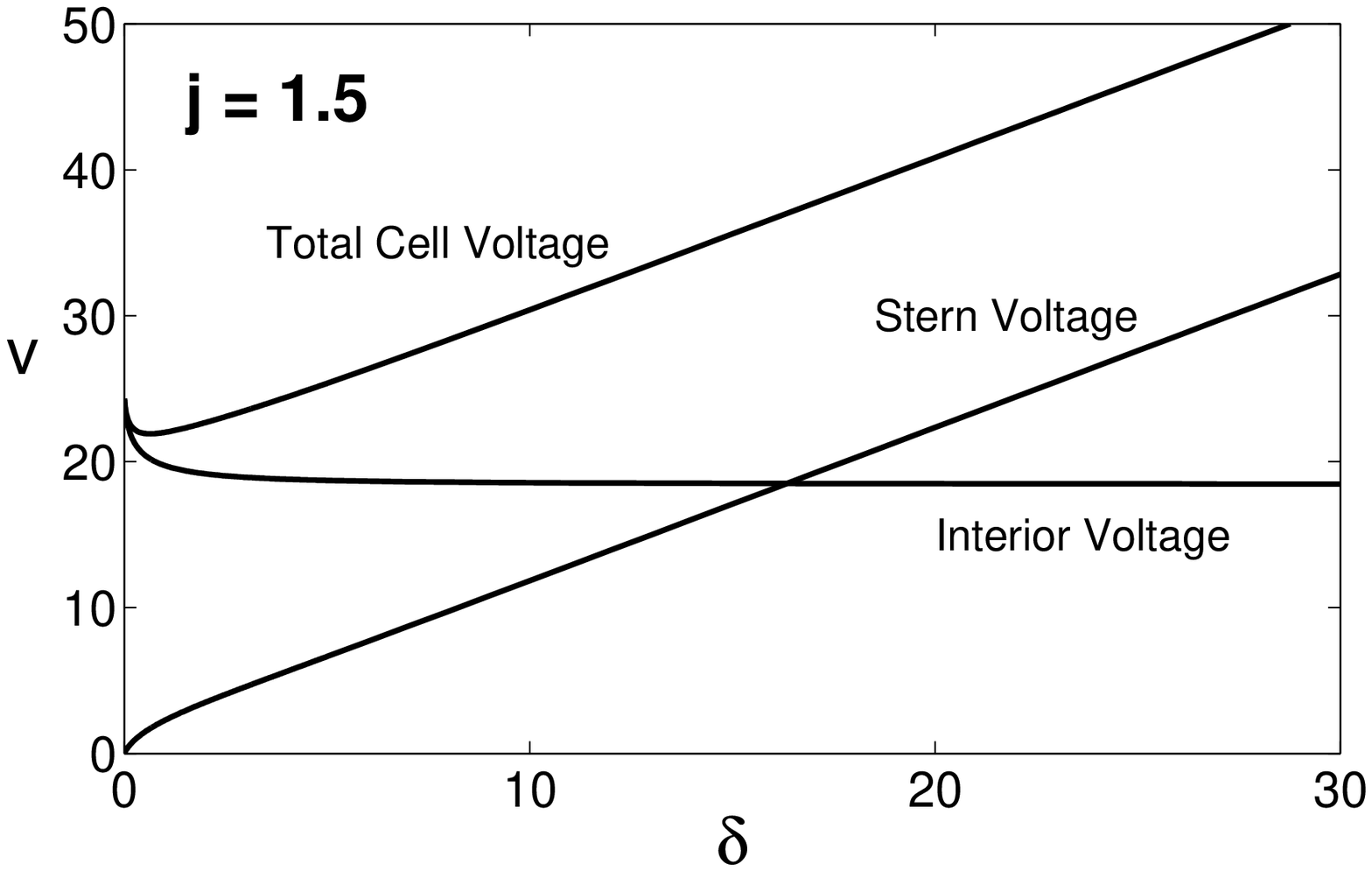}}
\begin{minipage}[h]{5in}
\caption{
\label{figure:stern_vs_interior}
These graphs break the total cell voltage into contributions from the
cell interior and the Stern layer as a function of $\delta$ for
$\epsilon = 0.01$, $k_c = 2$, and $\ir = 2$.  Note that at and above
the classical limiting current, the Stern layer voltage 
dominates the total cell voltage for large values of $\delta$.  }
\end{minipage}
\ec
\end{figure}

At high currents, another important effect of the Stern layer capacitance 
is that the total cell voltage becomes dominated by the potential drop
across the Stern layer at large $\delta$ values 
({\it i.e.} small capacitances).  This behavior is clearly illustrated
in Figure~\ref{figure:stern_vs_interior}.  Notice for currents
below the classical diffusion-limited current, the total cell voltage 
does not show a strong dependence on $\delta$.  However, for $\i > 1$,
the total cell voltage increases with $\delta$ -- the increase being
driven by the strong $\delta$ dependence of the Stern voltage.

\section{ Conclusion} 
In summary, we have studied classical problem of direct current in an
electrochemical cell, focusing on the exotic regime of high current
densities. A notable new feature of our study is the use of nonlinear
Butler-Volmer and Stern boundary conditions to model a thin film
passing a Faradaic current, as in a micro-battery.  We have derived
leading-order approximations for the fields at and above the
classical, diffusion-limited current, paying special attention to the
structure of the cathodic boundary layer, which must be present to
satisfy the reaction boundary conditions.  In our analysis of
superlimiting current, we have shown that the key feature of the bulk
space-charge layer is the depletion of anions. Our exact solution of
the leading-order problem in the space-charge region,
Eq.~(\ref{eq:Airysoln}), could thus also have relevance for Faradaic
conduction through very thin insulating films.  

Using the asymptotic approximations to the fields, we are able to
derive a current-voltage relation,
Eq.~(\ref{eq:v_vs_i_above_lim_cur}), which compares well with
numerical results, far beyond the limiting current. Combined with the
analogous formulae in the companion paper~\cite{part1}, which hold
below the limiting current, we have essentially analyzed the full
range of the current-voltage relation. These
results could be useful in interpreting experimental data, e.g. on
the internal resistance of thin-film microbatteries.  

A general conclusion of this study is that boundary conditions
strongly affect the solution. For example, the Stern-layer
capacitance, often ignored in theoretical analysis, plays an important
role in determining the qualitative structure of the cell near the
cathode, as well as the total cell voltage.  The nonlinear boundary
conditions for Butler-Volmer reaction kinetics also profoundly affect
charge distribution and current-voltage relation, compared to the
ubiquitous case of Dirichlet boundary conditions.  The latter rely on
the asumption of surface equilibrium, which is of questionable
validity at very large currents.

We leave the reader with a word of caution. The results presented here
are valid mathematical solutions of standard model equations, but
their physical relevance should be met with some skepticism under
extreme conditions, such as superlimiting current. For example, the
PNP equations are meant to describe infinitely dilute solutions in
relatively small electric
fields~\cite{newman_book,delahay_book,bard_book}. Even for
quasi-equilibrium double layers, their validity is not so clear when
the zeta potential greatly exceeds the thermal voltage, because co-ion
concentrations may exceed the physical limit required by discreteness
(accounting also for solvation shells) and counter-ion concentations
may become small enough to violate the continuum assumption. Large
electric fields can cause the permittivity to vary, by some estimates
up to a factor ten, as solvent dipoles become aligned. Including such
effects, however, introduces further {\it ad hoc} parameters into
the model, which may be difficult to infer from experimental data.

\section{Acknowledgments} 
This work was supported in part by the MRSEC Program of the National
Science Foundation under award number DMR 02-13282 and in part by the
Department of Energy through the Computational Science Graduate
Fellowship (CSGF) Program provided under grant number
DE-FG02-97ER25308.  The authors thank M. Brenner,
J. Choi, and B. Kim for helpful discussions.

\end{document}